\documentclass{article}
\usepackage{graphicx} 
\usepackage[a4paper, total={6in, 8in}]{geometry}

\usepackage[style=authoryear, sorting=nyt, url=false, eprint=false, maxcitenames=2, pluralothers=true, uniquename=false, natbib=true]{biblatex}

\DeclareCiteCommand{\citeyear}
    {}
    {\bibhyperref{\printdate}}
    {\multicitedelim}
    {}

\usepackage{dirtytalk}
\usepackage{colortbl}
\usepackage[labelfont=bf]{caption}
\counterwithin{figure}{section}

\usepackage{amsmath}
\numberwithin{equation}{section}
\usepackage{bbm}
\usepackage{amssymb}
\usepackage{mathtools}

\usepackage[usenames,dvipsnames]{xcolor}

\usepackage{hyperref}
\hypersetup{colorlinks = true, linkcolor = Mahogany, citecolor = CornflowerBlue, urlcolor = Periwinkle}

\allowdisplaybreaks

\usepackage{listings}

\lstdefinestyle{mystyle}{
    commentstyle=\color{OliveGreen},
    keywordstyle=\color{Magenta},
    numberstyle=\tiny\color{RoyalBlue},
    stringstyle=\color{BrickRed},
    basicstyle=\ttfamily\footnotesize,
    breakatwhitespace=false,         
    breaklines=true,                 
    captionpos=t,                    
    keepspaces=true,                 
    numbers=left,                    
    numbersep=15pt,                  
    showspaces=false,                
    showstringspaces=false,
    showtabs=false,                  
    tabsize=2
}

\title{Approaches to biological species delimitation based on genetic and spatial dissimilarity}
\author{Gabriele d'Angella\thanks{Alma Mater Studiorum - University of Bologna, Via Belle Arti, 41 - 40126 Bologna (BO), Italy.\hfill E-mail address: gabriele.dangella2@unibo.it}, Christian M. Hennig\thanks{Alma Mater Studiorum - University of Bologna, Via Belle Arti, 41 - 40126 Bologna (BO), Italy.\hfill E-mail address: christian.hennig@unibo.it}}
\date{December 12, 2025}

\begin{document}

\maketitle

\begin{abstract}
The delimitation of biological species, i.e., deciding which individuals belong to the same species and whether and how many different species are represented in a dataset, is key to the conservation of biodiversity. Much existing work uses only genetic data for species delimitation, often employing some kind of cluster analysis. This can be misleading, because geographically distant groups of individuals can be genetically quite different even if they belong to the same species. We investigate the problem of testing whether two potentially separated groups of individuals can belong to a single species or not, based on genetic and spatial data. Already existing methods such as the partial Mantel test and jackknife-based distance-distance regression are considered. New approaches, i.e., an adaptation of a mixed effects model, a bootstrap approach, and a jackknife version of partial Mantel, are proposed. All these methods address the issue that distance data violate the independence assumption for standard inference regarding correlation and regression. A standard linear regression is also considered. The approaches are compared on simulated meta-populations generated with the software packages SLiM and GSpace that can simulate spatially explicit genetic data at an individual level.
Simulations show that the new jackknife version of the partial Mantel test provides a good compromise between power and respecting the nominal type I error rate. Mixed-effects models have larger power than jackknife-based methods, but in some situations they display type I error rates above the significance level. 
An application on brassy ringlets concludes the paper.~
\end{abstract}

\section{Introduction}
\label{ch:Intro}

For the delimitation of biological species, empirical data is used to determine which groups of individual organisms constitute different populations of a single species and which constitute different species \citep{rannala2020}. Species delimitation is crucial for the preservation of biodiversity and has applications in several areas, such as ecology and medicine \citep{burbrink2021}. The empirical data employed to delimit species can be molecular (see, e.g., \cite{rannala2020} for a review), morphological \citep{gratton2016}, behavioural \citep{scapini2002}, ecological \citep{raxworthy2007, rissler2007}. There are also integrative approaches using different types of data \citep{edwards2014} and methods \citep{carstens2013}. 

Spatial information is key for this task, as witnessed by the increase in publications in the field of landscape genetics \citep{storfer2010}, which combines population genetics and landscape ecology \citep{balkenholCH1}. Neglecting geographical information when delimiting species can lead to misassessment of the genetic structure in the data \citep{frantz2009}. This can particularly happen in the presence of spatial patterns of genetic differentiation, such as isolation by (geographical) distance \citep[IBD;][]{ishida2009}: ignoring spatial information, individuals may be wrongly assigned to different species because their genetic dissimilarity tends to increase with geographical separation \citep{bradburd2018}, violating the often involved assumption of random mating within the population.

A way to include spatial information in molecular species delimitation routines is to study the relationship between genetic dissimilarity and geographical distance. The investigation of this relationship has a long tradition in the population genetics literature, where it was pursued to study migration models \citep{kimura64} or estimate demographic parameters \citep{slatkin93, rousset97, clarke2002}.
While isolation by distance assumes that the genetic dissimilarity between two individuals simply increases with a plain geographical distance (Euclidean distance, or the geodesic distance based on latitude and longitude), 
isolation by resistance \citep[IBR;][]{mcrae2006} takes into account landscape features such as rivers and heights: this translates to developments like the least-cost path approach \citep{adriaensen2003}, the circuit-based framework \citep{mcrae2008} or the least-cost transect analysis \citep{vanstrien2012}, in which a more sophisticated \say{landscape} dissimilarity measure is used. 
We will work with the Euclidean distance as geographical distance here, but all the considered methods also work with other geographical distances and dissimilarities.

\begin{figure}[tb] 
    \centering
    \includegraphics[width=0.49\textwidth]{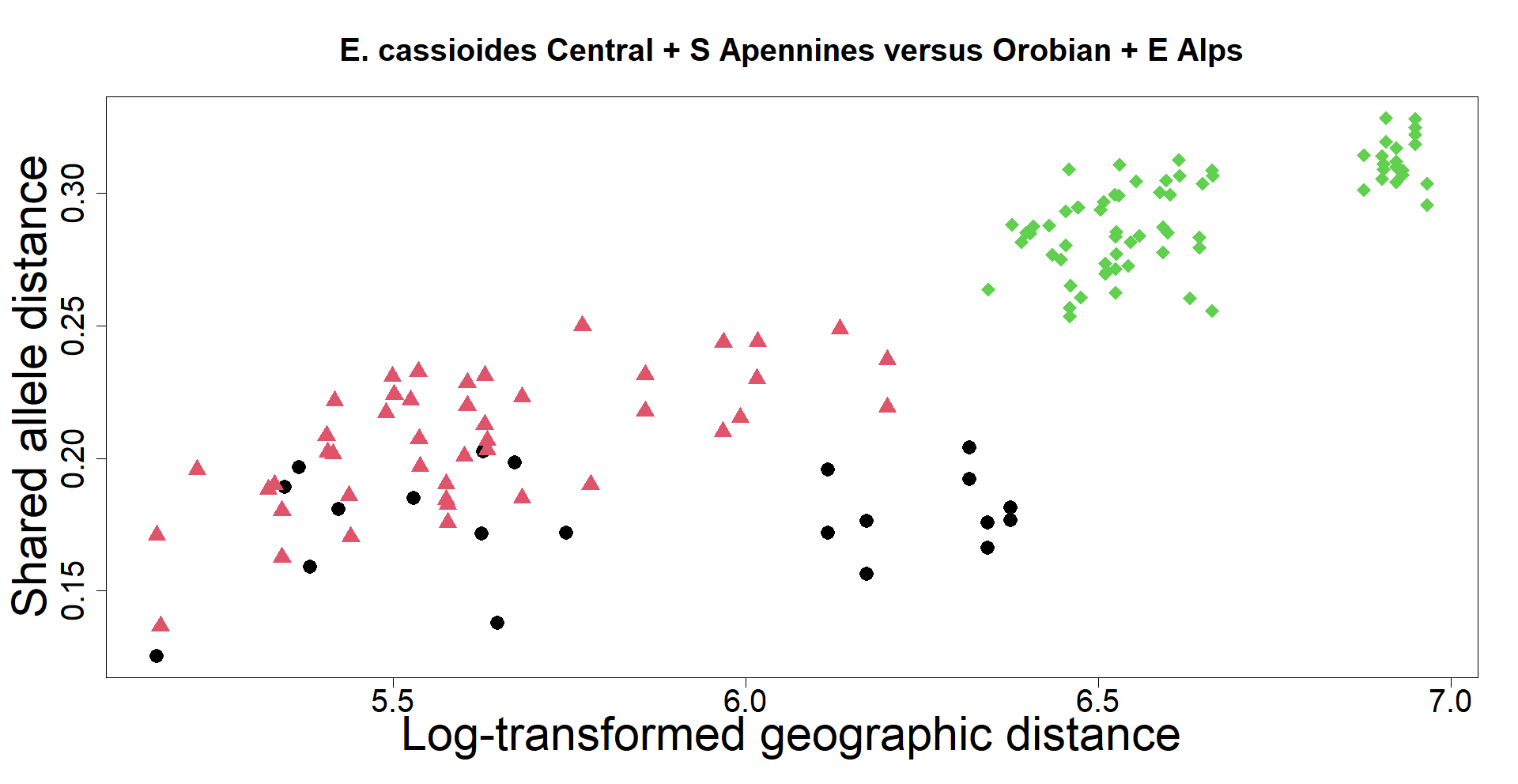}
    \includegraphics[width=0.49\textwidth]{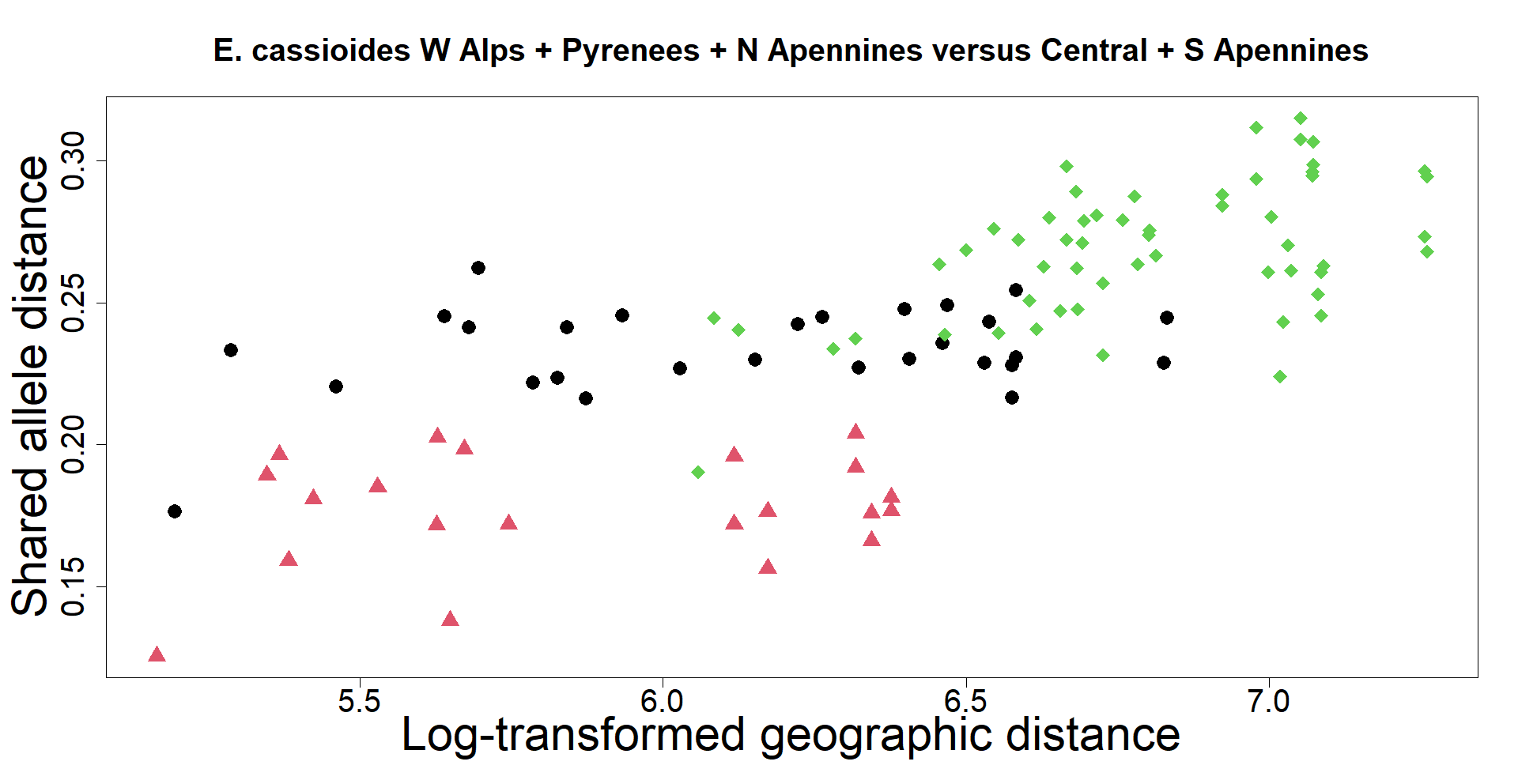}
    \includegraphics[width=0.49\textwidth]{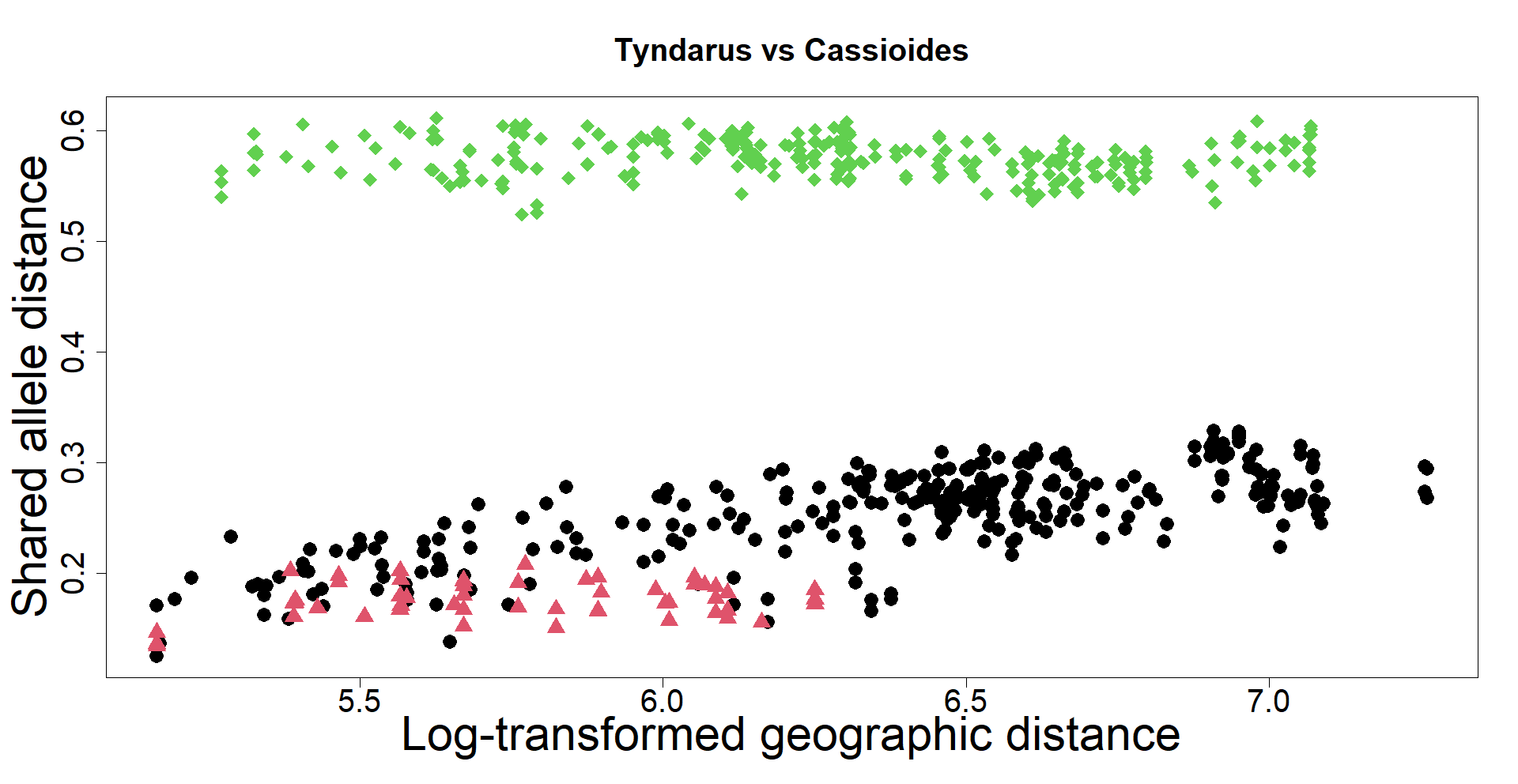}
    \caption{Log-transformed geographical distances vs. Shared allele distances for three pairs of groups from the brassy ringlets data (upper left side \textit{E. Cassioides central + S. Apennines} vs. \textit{Orobian + E. Alps}; upper right side \textit{E. Cassioides W. Alps, Pyrenees + N. Apennines} vs. \textit{central + S. Apennines}; lower plot \textit{E. Tyndarus} vs. \textit{E. Cassioides}), see Section \ref{ch:brassy}, for which conspecificity is of interest. The black circles (first group) and red triangles (second group) show distances between pairs of individuals belonging to the same group. The green diamonds show the distances between two individuals belonging to different groups.}
    \label{fig:brassydd}
\end{figure}

Consider a setup with two groups of individuals to be tested for conspecificity, i.e., being from the same species. Inference is based on checking whether the relationship between genetic and geographical dissimilarity differs between pairs of individuals in the same group and pairs in different groups (see Sections \ref{ch:data} and \ref{ch:brassy} for details). Each of the three panels in Figure \ref{fig:brassydd} shows genetic and geographical dissimilarities of two groups for which a test for conspecificity is of interest. Distances within the two groups are black circles and red triangles, distances between the groups are green diamonds. The plots show some (albeit weak) tendency that larger genetic distance comes with larger geographical distance, also within groups.  On the upper left side, genetic (``shared allele'') distances between groups seem slightly higher on average than genetic distances within groups, but also the geographical distances tend to be higher, and just from looking at the data it is not clear cut whether larger genetic distances between groups can be explained by the geographical distances only (in which case there is no reason to consider the two groups as different species), even less so on the upper right side. In the lower plot, it is clear that genetic distances between groups are much larger than they could be expected to be in case the two groups belonged to the same species. 

The impact of grouping on the genetic dissimilarity can be quantified controlling for the effect of geographical distance. \cite{medrano2014} used a permutation-based partial Mantel test \citep[PMT;][]{SLS86} to assess the significance of the partial correlation coefficient between genetic and grouping dissimilarities given the geographical distance, where the grouping dissimilarity is defined as 0 if a pair of individuals is in the same group and 1 otherwise.
\cite{hh20} suggested to jackknife test for whether a regression fitted on the within-group distances can also explain the between-groups distances.
\cite{clarke2002} employed individual random effects in order to model the dependence between dissimilarities of the same kind belonging to the same individual. This approach can be extended and adapted to the IBD problem by testing for an effect of the grouping dissimilarity as we do in the present paper. As further new approaches, we consider partial Mantel tests using jackknife or bootstrap instead of permutations. All these techniques have in common that they use statistics that were originally devised for situations with i.i.d. (independently identically distributed) data or residuals. Dissimilarities involving the same individual are certainly not independent, but the methods account for dependence by using permutations or jackknife at the individual rather than the dissimilarity level, or by modelling dependence using a random effect.
For exploring how much of a difference taking into account the dependence between dissimilarities actually makes, we also consider a multiple regression with genetic dissimilarities as response and geographical and grouping dissimilarities as explanatory variables, ignoring dependence between dissimilarities. 

There exists only anecdotal evidence of the performance of the methods in \cite{medrano2014}, \cite{spriggs2019} and \cite{hh20}; they have not been systematically assessed from a statistical perspective.
The study presented here consists in a systematic comparison of the type I error rate and power of the aforementioned methods based on simulating spatially-explicit genetic datasets generated by the simulators GSpace \citep{gspace2021} and SLiM \citep{slim2023}. 

\begin{figure}[tb] 
    \centering
    \includegraphics[width=0.75\textwidth]{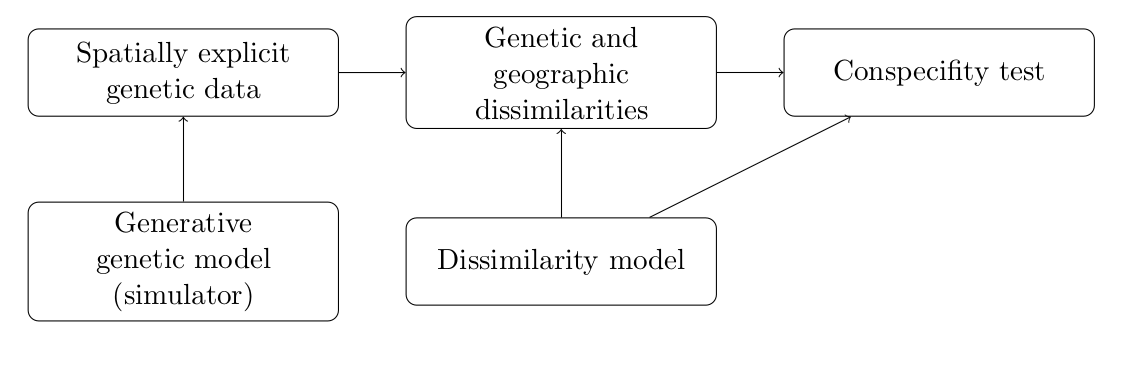}
    \caption{The conspecificity tests treated here are computed on genetic and geographical dissimilarities assuming models for general dissimilarities (i.e., not taking into account how exactly the dissimilarities came about). The dissimilarities are computed from the originally observed spatially explicit genetic data, which in our simulation study are simulated from generative genetic models.}
    \label{fig:geneticdisflow}
\end{figure}

A distinctive feature of the present study is that the genetic dissimilarities on which inference is based are much simpler than the data from which they are computed, see Section \ref{ch:data}. To our knowledge, due to complexity, currently no tests for conspecificity exist that operate directly on the genetic information and take into account geographical distance. GSpace and SLiM provide sophisticated models for the genetic data, but the inference does not use such models. Instead it is based on much simpler models for the dissimilarities without taking into account how these were computed from the original genetic data, see Figure \ref{fig:geneticdisflow}. Here we confront such inference with the more complex genetic models for exploring its statistical characteristics. Methods and results may be relevant also for other problems where regression between dissimilarities is of interest.

The species concept in biology is somewhat controversial \citep{hausdorf2009} and there is biological differentiation between populations at different levels: there are species that are more or less closely related, and there is differentiation also below the species level - see, e.g., \cite{dequeiroz2007}. Because of this, any result of the treated tests should not be taken as conclusive regarding conspecificity. The aim is just to formalise a key aspect of the information in the data. The tests developed here test for discontinuities in the gene flow. These may occur because different populations are actually different species, but could also be caused, for example, by geographical barriers.


This paper is organized as follows: the data and distances are introduced in Section \ref{ch:data}, then all methods are discussed in Section \ref{ch:methods}. The two simulators used in this study and the results obtained with them are discussed in Section \ref{ch:simulations}. Section \ref{ch:brassy} presents an application to the brassy ringlets data examined by \cite{gratton2016}. Section \ref{ch:conclusion} concludes the paper.

\section{Data and dissimilarities}\label{ch:data}

Spatially explicit genetic data consists of individuals carrying information about their location and genetic make-up. Methods will be applied on individuals from two groups, with known membership. Hence, two columns will correspond to the unit's coordinates (northings and eastings, latitude and longitude, etc.), one column will report the group labels (either group $1$ or $2$) and the other $P$ columns will be loci (locations on the DNA). Individual-level codominant data, such as SNPs or microsatellites, with diploid genotypes will be considered \citep{balkenholCH3}: this means that each locus will contain two alleles. Following \cite{prabclus} we represent alleles by single characters although elsewhere in the literature more elaborate coding is used  \citep{genepop}.
The resulting $n\times(P+3)$ data frame will be denoted by
\[ \mathbf{Z} = \begin{pmatrix}
        \mathbf{z}_1 \\
        \vdots \\
        \mathbf{z}_n \\
        \end{pmatrix}
        = \begin{pmatrix} 
        z_1^{(x)} & z_1^{(y)} & z_1^c & Z_1^1 & \cdots & Z_1^P \\
        z_2^{(x)} & z_2^{(y)} & z_2^c & Z_2^1 & \cdots & Z_2^P \\
        \vdots & \vdots & \vdots & \vdots & \ddots & \vdots \\
        z_n^{(x)} & z_n^{(y)} & z_n^c & Z_n^1 & \cdots & Z_n^P \\
        \end{pmatrix} \,,
\]
where each observed locus $Z_i^p$, $p=1,\ldots,P$, is a set of characters, like $\{A, B\}$ for heterozygous loci (\say{BA}) or $\{B\}$ for homozygous ones (\say{BB}); also note that alleles are arranged in lexicographical order in the sets because a meaningful order is not normally observable. Each $\mathbf{z}_i$ is a $1\times(P+3)$ vector representing the $i^{th}$ individual. $z_i^{(x)}, z_i^{(y)}$ are geographical coordinates, and $z_i^c\in \{1, 2\}$ is a group indicator, where groups 1 and 2 represent the candidate species to be tested for conspecificity.

In this study, the Euclidean distance will be employed as geographical distance (subscript $x$):
\begin{equation*} 
  d_x(\mathbf{z}_i, \mathbf{z}_j) = \sqrt{\left(z_i^{(x)} - z_j^{(x)}\right)^2 + \left(z_i^{(y)} - z_j^{(y)}\right)^2}\,.
\end{equation*}
As genetic dissimilarity (subscript $y$), the shared allele dissimilarity \citep{bowcock94} will be used:
\begin{equation}
\label{shald}
  d_y(\mathbf{z}_i, \mathbf{z}_j) = 1 - \frac{1}{2P}\sum_{p=1}^P \left|Z_i^p \cap Z_j^p \right| \cdot \left[1 + \mathbbm{1}\left(|Z_i^p| + |Z_j^p| = 2\right)\right],
\end{equation}
where $\mathbbm{1}(\text{condition})=1$ if the condition is true and zero otherwise. In real data occasionally there is missing data (missing loci). In this case $d_y$ just averages over the loci that are non-missing in both $\mathbf{z}_i$ and $\mathbf{z}_j$. If there are no missing values, the shared allele dissimilarity is actually a distance (see proof in Supplementary Material), but missing values can cause a violation of the triangle inequality.

It is easy to see that, the larger $P$, the finer is the quantification of the genetic dissimilarity between two species, as more sites are available for the comparison of two individuals' genetic information.

Let $n_j = |\{i: z_i^c = j\}|$ be the number of individuals belonging to group $j=1,2.$ In practice, this grouping information can be based on morphological, behavioural or even spatial grounds or can simply represent the researcher's hypothesis. The number of geographical distances in the dataset amounts to:
\begin{equation*}
  \frac{1}{2}(n_1 + n_2)(n_1 + n_2 - 1) = \underbrace{\frac{1}{2}n_1(n_1 - 1)}_{\text{within group 1}} + \underbrace{\frac{1}{2}n_2(n_2 - 1)}_{\text{within group 2}} + \underbrace{n_1n_2}_{\text{between groups}} \,,
\end{equation*}
to be stored in the following $n\times n$ block matrix, with $n = n_1 + n_2$,
\begin{align*}
  \mathbf{D}_x 
  &= \left(\begin{array}{c|c}
      \mathbf{D}_x^{11} & \mathbf{D}_x^{12} \\
     \hline
      \mathbf{D}_x^{21} & \mathbf{D}_x^{22} \\
      \end{array}
    \right) \\ 
   &=
    \resizebox{0.9\textwidth}{!}{ $
    \left(\begin{array}{cccc|ccc}
    0 & d_x(\mathbf{z}_1, \mathbf{z}_2) & \cdots & d_x(\mathbf{z}_1, \mathbf{z}_{n_1}) & d_x(\mathbf{z}_1, \mathbf{z}_{n_1 + 1}) & \cdots & d_x(\mathbf{z}_1, \mathbf{z}_{n_1 + n_2}) \\
      d_x(\mathbf{z}_2, \mathbf{z}_1) & 0 & \cdots & d_x(\mathbf{z}_2, \mathbf{z}_{n_1}) & d_x(\mathbf{z}_2, \mathbf{z}_{n_1 + 1}) & \cdots & d_x(\mathbf{z}_2, \mathbf{z}_{n_1 + n_2}) \\
      \vdots & \vdots & \ddots & \vdots & \vdots & \ddots & \vdots \\
      d_x(\mathbf{z}_{n_1}, \mathbf{z}_1) & d_x(\mathbf{z}_{n_1}, \mathbf{z}_2) & \cdots & 0 & d_x(\mathbf{z}_{n_1}, \mathbf{z}_{n_1 + 1}) & \cdots & d_x(\mathbf{z}_{n_1}, \mathbf{z}_{n_1 + n_2}) \\
      \hline
      d_x(\mathbf{z}_{n_1+1}, \mathbf{z}_1) & d_x(\mathbf{z}_{n_1+1}, \mathbf{z}_2) & \cdots & d_x(\mathbf{z}_{n_1+1}, \mathbf{z}_{n_1}) & 0 & \cdots & d_x(\mathbf{z}_{n_1}, \mathbf{z}_{n_1 + n_2}) \\
      \vdots & \vdots & \ddots & \vdots & \vdots & \ddots & \vdots \\
      d_x(\mathbf{z}_{n_1+n_2}, \mathbf{z}_1) & d_x(\mathbf{z}_{n_1+n_2}, \mathbf{z}_2) & \cdots & d_x(\mathbf{z}_{n_1+n_2}, \mathbf{z}_{n_1}) & d_x(\mathbf{z}_{n_1+n_2}, \mathbf{z}_{n_1+1}) & \cdots & 0 \\
    \end{array} \right) $} \,, 
\end{align*}
where matrix $\mathbf{D}_x^{11}$ stores the distances among the observations belonging to group 1, $\mathbf{D}_x^{22}$ those within group 2 and $\mathbf{D}_x^{12}=(\mathbf{D}_x^{21})^\intercal$ those among individuals of different groups. $\mathbf{D}_x$ carries redundant information: it is sufficient to work with the lower triangular matrix $\{ d_x(\mathbf{z}_r, \mathbf{z}_c) \}_{r>c}$.
Analogously, the $n\times n$ matrices $\mathbf{D}_y$ and $\mathbf{D}_g$ store the genetic and grouping dissimilarities, with $d_g(\mathbf{z}_i, \mathbf{z}_j) = \mathbbm{1}(z_i^c \ne z_j^c)$.

We will later use the sets of all index pairs referring to within-group and between-group dissimilarities, respectively:
\[
W = \Bigl\{ (r, c) | \, c < r \leq n_1 \lor n_1 < c < r \leq n \Bigl\},\
B = \Bigl\{ (r, c) | \, n\geq r > n_1 \land c \leq n_1 \Bigl\}.
\]
In order to improve the linearity between geographical and genetic dissimilarities, it is often advantageous to use log-transformed geographical distances as is done for example in \cite{vekemanshardy2004,rousset97,hh20}. Zero geographical distances can occur if two individuals are observed in the same location. Therefore, following \cite{hh20}, the following transformation is considered:
\begin{equation}
\label{eq:logx}
f(d_x(\mathbf{z}_r, \mathbf{z}_c)) = \ln(d_x(\mathbf{z}_r, \mathbf{z}_c) + F_x^{-1}(0.25))\,,
\end{equation}
where $F_x$ is the empirical cdf of all geographical distances in the dataset, and $F_x^{-1}(\alpha)$ is the corrsponding $\alpha$-quantile.
In our comparative study we will  consider both untransformed and log-transformed $\mathbf{D}_x$. In order to keep notation light, all methods will be described using untransformed geographical distances.

Although the shared allele distance is used here, the discussed methods are based on models for dissimilarities that do not rely on the specific dissimilarity. The methods can therefore also be applied to other dissimilarities. In conspecificity testing, sometimes data come at population level with genetic distances between populations rather than individuals, e.g., in \cite{clarke2002} and one example in \cite{hh20}.
\section{Methods}\label{ch:methods}
The methodologies presented here use the information on the relationship between the distances in $\mathbf{D}_y$ and $\mathbf{D}_x$ with the aim of testing a conspecificity presumption encoded in $\mathbf{D}_g$. In the presence of isolation by distance behaviour (positive association between genetic and geographical dissimilarities), two groups of individuals belonging to the same species might display a certain degree of genetic structure that is explained by their geographical separation. If the genetic dissimilarities are too large to be compatible with the geographical separation between the two groups, this will constitute evidence for lineage separation, i.e., for distinctness.
As \cite{hh20}
write, \say{\textit{it is often difficult to assess whether observed differences between allopatric meta-populations
would be sufficient to prevent the fusion of these meta-populations upon contact.}} In these situations,
non-spatial models (to whom the putative grouping is often ascribed in practical applications) may be biased, and IBD patterns should be taken into account \citep{meirmans2012}.

The computation of dissimilarities implies information loss: the complex biological mechanisms \citep[e.g., dispersal, see][]{cayuela2018} that act on the allele frequencies of the two investigated putative species have an indirect effect on the relationship between genetic and geographical dissimilarities, which can be nonlinear \citep{hutchison1999}. The methods discussed here do not attempt to model such evolutionary processes, but rather work at the dissimilarity level, where the information from the $P$ loci is summarized.
The conspecificity null hypothesis is operationalised by these methods as having the same trend in the relation between genetic and geographical dissimilarities within groups and between groups. For the alternative hypothesis, genetic dissimilarities would be expected to be larger between groups than within groups when adjusted for geographical distances. The methods are based on linear regression and correlation, i.e., they are based on a linearity assumption. Note however that it can normally be expected with enough data that a zero correlation or regression slope can also be rejected if the relation is nonlinear but monotonic. Therefore the methods can be used also to detect nonlinear monotonic deviations from the null hypothesis, even though linearity would be the ideal condition. Incidentally, for Euclidean data, \cite{szekely2007} even show that independence is equivalent to a \say{distance correlation}, closely related to what is considered here, being zero. Nonlinearity occurs in some real datasets for example because of saturation effects, i.e., genetic distances reaching the maximum possible value; the Supplementary Material shows examples.
Furthermore, the methods treated here do not require the triangle inequality, and monotonic transformations of dissimilarities can also be used.

All methods considered here are heuristic with weak (if any) theoretical justification for the given situation. Some of the methods have been explicitly criticized for the violation of model assumptions of existing theory, see below. To our knowledge, however, there are no alternatives with a stronger foundation, and the methods that are already published are used in practice. In general, statistical model assumptions can be seen as idealizations and are rarely (if at all) fulfilled in practice. This means that the violation of certain assumptions and the lack of theory applying to the specific situation does not automatically make a method invalid. The relevant question is whether certain issues with the data (such as violation of standard assumptions) have the potential to cause a method to produce seriously misleading results. For this reason an empirical investigation as done here should be worthwhile; we also give some remarks regarding the validity of the heuristics. More theoretical investigation is desirable, but it will probably be hard and is not the topic of this paper. Note in particular that our aim is not to advertise all involved methods or a specific one, but rather to explore strengths and weaknesses of all methods, and certainly all of them need to be applied cautiously.


In the following, the statistical methods involved in this comparative study are described.

\subsection{Regression on dissimilarities with jackknife testing}\label{sec:hh20}
\cite{hh20} proposed to regress the genetic dissimilarities on the log-transformed geographical distances trying to clarify whether the genetic structure found between the two candidate species can be compatible with their IBD behaviour.
To this end, a regression line based on the within-group dissimilarities (red and black observations in Figure \ref{fig:brassydd}) is compared with a regression line based on all dissimilarities. The null hypothesis of conspecificity is rejected if the between-groups dissimilarities (green in Figure \ref{fig:brassydd}) are systematically too large compared to what would be expected from the regression computed on the within-group dissimilarities. Dependence between dissimilarities is taken into account by running the test using a jackknife scheme that treats the individuals rather than the dissimilarities as observational units.

This approach is complicated by the fact that the test just mentioned relies on a single regression line being appropriate for the within-group dissimilarities in both groups. \cite{hh20} propose a test protocol where it is first tested whether this is the case ($H_{01}$). Then, depending on the result, either a null hypothesis of a joint regression for all dissimilarities is tested ($H_{02}$, corresponding to conspecificity), or, in case that $H_{01}$ is rejected, it is tested whether the between-groups distances are in line with at least one of the group-wise regressions of the within-group dissimilarities ($H_{03}$; in case that this is rejected, it is taken as evidence against conspecificity, whereas non-rejection is an ambiguous result that would need closer biological investigation).

The first of the three tests focuses on the relationship between genetic and geographical dissimilarities within the two groups, assuming the following linear relationship:
\begin{equation} \label{eq:hh20_h01}
  d_y(\mathbf{z}_r, \mathbf{z}_c) = \Biggl\{
  \begin{array}{ccccc}a_1 + b_1\{ d_x(\mathbf{z}_r, \mathbf{z}_c)   - \bar{d_{x}}^{W}\} + e(\mathbf{z}_r, \mathbf{z}_c) & & \text{with}\, c < r\leq n_1\\
  a_2 + b_2\{ d_x(\mathbf{z}_r, \mathbf{z}_c)   - \bar{d_{x}}^{W}\} + e(\mathbf{z}_r, \mathbf{z}_c) & & \text{with}\, n_1 < c < r
  \end{array}.
\end{equation}
$a_1$, $b_1$, $a_2$ and $b_2$ are estimated via least squares, and
\[ 
\bar{d_x}^{W} = \frac{1}{|W|} \sum_{r,c \in W} d_x(\mathbf{z}_r, \mathbf{z}_c)
\]
is the mean within-group geographical distance taken over both candidate species.
The errors $e$ in (\ref{eq:hh20_h01}) are assumed to have zero mean, but not to be independent. Only the genetic random variation of individuals is assumed to be independent, but not dissimilarities involving the same individual. 

The first test tests $H_{01}:\ a_1 = a_2$ and $b_1 = b_2$. It is tested against the two-sided alternative that $a_1 - a_2 \ne 0$ \textit{or} $b_1 - b_2 \ne 0$. Both of these are tested and combined using Bonferroni, i.e., multiplying the minimum of the two p-values by 2. 


In order to deal with the dependence between dissimilarities, \cite{hh20} use non-parametric jackknife (already suggested by \cite{clarke2002}) to obtain a measure of the variability of the estimates. Jackknifing \citep[][ch. 11]{efron1993} here consists in computing as many OLS estimates as the number of individuals involved in a given regression model (e.g., $n_1$ for group 1) by fitting it on the $n_1$ datasets obtained by removing one individual at a time. In this particular setup, the removal of one individual implies the removal of all the dissimilarities related to it, so each jackknife replicate of the OLS estimates for group 1 is based on $(n_1 - 1)(n_1-2)/2$ data points instead of $n_1(n_1-1)/2$. 

In jackknifing, so-called pseudovalues $u_i,\ i=1,\ldots,n,$ for a statistic $U$ computed on data ${\bf X}$ with $n$ observations are computed as $u_i=nU({\bf X})-(n-1)U({\bf X}_{(i)})$ where ${\bf X}_{(i)}$ has the $i^{th}$ observation left out. The variability of the difference between parameter estimates is quantified by pooling the within-group jackknife estimates of standard error \citep[][ch. 11]{efron1993} in order to run a Welch's t-test \citep{welch47}.

This principle is applied here to both the difference between intercepts and to the difference between slopes of the two within-group regressions, where the null hypothesis for Welch's t-test is that the expected difference is zero, see \cite{hh20} for more details. Jackknife testing is a heuristic idea that has a theoretical justification only in specific situations \citep{shaowu89}, the assumptions of which are not fulfilled here. Therefore, its characteristics have to be explored experimentally in all but the simplest situations, which is done here in Section \ref{ch:simulations}. Jackknifing individuals treats the individuals rather than the dissimilarities as independent units of the analysis, and can be expected to lead to a larger jackknife standard error, and therefore more conservative tests than jackknifing dissimilarities. This means in particular that the dependence between dissimilarities belonging to the same individual will not specifically invalidate the jackknife test (the t-test is not run on the dissimilarities but on the jackknife pseudovalues as originally proposed by Tukey, see \cite{miller74}) beyond the fact that it is not covered by the existing theory.

If $H_{01}$ is not rejected, a unique regression is fitted on all the within-group dissimilarities, regardless of the membership, because the IBD behaviour of the two candidate species looks compatible. In this situation, hypothesis $H_{02}$ is tested.
The following ordinary least squares model is fitted:
\begin{equation} \label{eq:hh20_h02wg}
  d_y(\mathbf{z}_r, \mathbf{z}_c) = a_* + b_*(d_x(\mathbf{z}_r, \mathbf{z}_c) - \bar{d_x}^{W}) + e(\mathbf{z}_r, \mathbf{z}_c),
\end{equation}
where $\mathbb{E}(e(\mathbf{z}_r, \mathbf{z}_c))=0$ and $r,c\in W$.
This fit will be compared with the following model, which is based on all the dissimilarities in the dataset (within and between-group), regardless of the grouping:
\begin{equation} \label{eq:hh20_h02all}
  d_y(\mathbf{z}_r, \mathbf{z}_c) = a + b(d_x(\mathbf{z}_r, \mathbf{z}_c) - \bar{d_x}^{W}) + e(\mathbf{z}_r, \mathbf{z}_c),
\end{equation}
where $c<r\leq n$ and $\mathbb{E}(e(\mathbf{z}_r, \mathbf{z}_c))=0$. Define
$\bar{d_x}^B = \frac{1}{|B|}\sum_{r,c\in B} d_x(\mathbf{z}_r, \mathbf{z}_c)$,
the average between-group geographical distance. $H_{02}:\ a=a^*$ and $b=b^*$ is then tested against the one-sided alternative 
\begin{equation} \label{eq:hh20_h02alt}
a + b(\bar{d_x}^B - \bar{d_x}^{W}) > a_* + b_*(\bar{d_x}^B - \bar{d_x}^{W}), 
\end{equation}
i.e., genetic dissimilarities predicted at $\bar{d_x}^B$ by all dissimilarities combined are systematically larger than predicted by within-group dissimilarities only. The statistic on which 
jackknife testing is based is
\begin{equation}
\label{eq:hh20_h02t}
    \hat{a} + \hat{b}(\bar{d_x}^B - \bar{d_x}^{W}) - \hat{a_*} - \hat{b_*}(\bar{d_x}^B - \bar{d_x}^{W}),
\end{equation}
where $\hat{a}$, $\hat{b}$, $\hat{a_*}$ and $\hat{b_*}$ are the corresponding OLS estimates.

If $H_{01}$ is rejected, the IBD behaviour of the two candidate species cannot be described by a unique model and model (\ref{eq:hh20_h01}) is adopted. In this situation, $H_{03}$ is tested, that is, the compatibility of IBD behaviour and genetic structure is checked for each group separately. Two models similar to (\ref{eq:hh20_h02all}) are set up, each one comprising within-group dissimilarities from one of the groups only, together with the between-group dissimilarities, and two jackknife tests are run on statistics analogous to (\ref{eq:hh20_h02t}). $H_{03}:\ a_j=a^*_j$ and $b_j=b^*_j$ for at least one of $j=1,2$ is tested, where $a_j, b_j$ refer to regressions based on dissimilarities within group $j$ only, and  $a_j^*, b_j^*$ refer to regressions based on all dissimilarities involving a member of group $j$. The alternative is defined by analogy to (\ref{eq:hh20_h02alt}). The test rejects $H_{03}$ if the maximum of the p-values for the two tests regarding groups $j=1, 2$ is too small. Note that jackknifing these tests involves two different kinds of pseudovalues. For example consider the test regarding group 1. Members of group 1 are involved in dissimilarities between other members of group 1 and group 2, whereas members of group 2 are only involved in dissimilarities to members of group 1. This is accounted for by the computation of the jackknife estimate of the standard error to be used for the t-test, see \cite{hh20} and the documentation of the \texttt{prabclus R}-implementation \citep{prabclus} for details.

A rejection to the test for either $H_{02}$ or $H_{03}$ constitutes evidence against the null hypothesis of conspecificity, suggesting that the relationship between genetic and geographical dissimilarities displayed by the two meta-populations cannot explain their genetic differences and they might thus represent two separated lineages.

\subsection{The partial Mantel test}\label{thePMT}

The null hypothesis of the simple Mantel test states that \say{\textit{the distances among objects in matrix $\mathbf{D}_y$ are not (linearly or monotonically) related to the corresponding distances in $\mathbf{D}_x$}} \citep[][p. 600]{legendre2012}. The original test statistic by \cite{mantel67} was a cross-product of the vectors of dissimilarities,
\begin{equation*}
    \sum_{c<r\leq n} d_y(\mathbf{z}_r, \mathbf{z}_c) \cdot d_x(\mathbf{z}_r, \mathbf{z}_c) \, ,
\end{equation*}
the standardized version of which corresponds to the sample correlation coefficient between the vectors of dissimilarities:
\begin{align}
\label{eq:simpleMantel}
    r(\mathbf{D}_y, \mathbf{D}_x) = \frac{\sum_{c<r\leq n} (d_y(\mathbf{z}_r, \mathbf{z}_c) - \bar{d}_y)(d_x(\mathbf{z}_r, \mathbf{z}_c) - \bar{d}_x)}{\sqrt{\sum_{c<r\leq n} (d_y(\mathbf{z}_r, \mathbf{z}_c) - \bar{d}_y)^2 \sum_{c<r\leq n} (d_x(\mathbf{z}_r, \mathbf{z}_c) - \bar{d}_x)^2}},
\end{align}
where $\bar{d}_y = \frac{1}{w}\sum_{c<r\leq n} d_y(\mathbf{z}_r, \mathbf{z}_c)$ is the overall average genetic dissimilarity and $\bar{d}_x = \frac{1}{w}\sum_{c<r\leq n} d_x(\mathbf{z}_r, \mathbf{z}_c)$ is the overall average geographical distance, with $w=n(n-1)/2$.

Partial Mantel tests were proposed by \cite{SLS86} and are based on a partial correlation coefficient here defined as
\begin{equation}
\label{eq:partialMantel}
    r(\mathbf{D}_y, \mathbf{D}_g | \mathbf{D}_x) = \frac{r(\mathbf{D}_y, \mathbf{D}_g) - r(\mathbf{D}_y, \mathbf{D}_x) r(\mathbf{D}_g, \mathbf{D}_x)}{\sqrt{(1 - r(\mathbf{D}_y, \mathbf{D}_x)^2) (1 - r(\mathbf{D}_g, \mathbf{D}_x)^2)}}\,.
\end{equation}
(\ref{eq:partialMantel}) quantifies the correlation between the genetic dissimilarities and the grouping distances after having accounted for the geographical distances.  
\cite{medrano2014} tested the null hypothesis that 
$\rho(\mathbf{D}_y, \mathbf{D}_g | \mathbf{D}_x)=0$, where $\rho$ is the true underlying partial correlation in the population in order to ascribe the genetic structure found in two subgroups of trumpet daffodils to their lineage separation. The rejection of such hypothesis led them to maintain that the IBD behaviour displayed by the groups was not sufficient to explain the genetic dissimilarity found between the groups and that these should therefore not be considered conspecific. In Figure \ref{fig:brassydd} the null hypothesis means that conditionally on geographical distances between-groups genetic dissimilarities (i.e., green) are not systematically larger than within-group ones (i.e., red or black).

Hypothesis testing is usually carried out by means of permutations. 
\cite{legendre2000} carried out empirical comparisons of four permutation strategies for partial Mantel tests. His first strategy, the one used in this study, consists in permuting just one of the three dissimilarity matrices and recomputing the partial correlation coefficient a large number of times. 
The default number of permutations in the \texttt{ecodist} package by \cite{goslee2007}, which was used in this study, is 1000. For each of these $1000$ iterations, rows and corresponding columns in matrix $\mathbf{D}_y$ are permuted to yield $\mathbf{D}^*_y$, which implies the modification of $r(\mathbf{D}^*_y, \mathbf{D}_g)$ and $r(\mathbf{D}^*_y, \mathbf{D}_x)$ to be included in (\ref{eq:partialMantel}). If the two groups are separated species, the partial correlation between genetic and group dissimilarities should be positive (larger genetic dissimilarity between groups). Therefore    
a one-sided test is carried out, and the associated $p$-value is equal to the share of $r(\mathbf{D}^*_y, \mathbf{D}_g | \mathbf{D}_x)$ permutation replicates that are at least as large as the original value $r(\mathbf{D}_y, \mathbf{D}_g | \mathbf{D}_x)$. \cite{legendre2000} remarked that this permutation strategy may lead to inflated type-I error if outlying dissimilarity values are present in the data, whereas skewness in the dissimilarities distribution should not represent an issue. Another difficulty is that the entries in $\mathbf{D}_g$ are binary, and Pearson's correlation is originally defined for continuous entries. It has been applied as point-biserial correlation also to binary vectors, and the assessment of significance by the permutation principle accounting for dependence of entries of $\mathbf{D}_g$ in the same row or column is arguably not invalidated by binary data. Mantel tests can be run with other correlation  measures, see \cite{dietz83}, but it is not obvious why this should lead to improvements.

\paragraph{Testing with jackknife}
Significance in partial Mantel tests is typically assessed via permutations. This, however, might introduce a distortion. Permuting $\mathbf{D}_y$ while keeping $\mathbf{D}_g, \mathbf{D}_x$ fixed generates data for which $\rho(\mathbf{D}_y, \mathbf{D}_g | \mathbf{D}_x)=0$ as prescribed by the null hypothesis. However, on top of that, the permuted $\mathbf{D}_y$ will be independent of both $\mathbf{D}_g$ and $\mathbf{D}_x$, which may be inappropriate in a real situation. Other permutation schemes as listed in \cite{legendre2000} also come with potentially unrealistic implicit structural assumptions. Partial Mantel tests have been controversially discussed with mixed empirical results in various situations, see, e.g., \cite{guillot2013, legendre2015}.

The potential distortion from permutation can be prevented by jackknifing the partial correlation (\ref{eq:partialMantel}), leaving one individual out at a time, and then generate pseudovalues and run a t-test as explained in Section \ref{sec:hh20}.

\paragraph{Testing with bootstrap}
Another option to assess the variability of the partial correlation coefficient $r(\mathbf{D}_y, \mathbf{D}_g | \mathbf{D}_x)$ is by resampling $n$ individuals with replacement, generating nonparametric bootstrap samples. This idea was discouraged in \cite{clarke2002} and \cite{hh20} because, whenever two identical individuals are sampled more than once, the associated dissimilarities will be equal to zero, generating bootstrap samples that in most cases tend to display a larger proportion of zero dissimilarities with respect to the original data. However, to date, no systematic study has demonstrated the performance of nonparametric bootstrap for species delimitation tasks.

A seminal reference for this technique is \cite{efron1993}. We use the  \textit{bias-corrected} (BC) bootstrap confidence intervals here, defined and motivated in \cite[][ch. 22.5]{efron1993}. The null hypothesis of conspecificity is rejected in the lower bound of a $(1-2\alpha)$ bootstrap confidence interval ($\alpha$-quantile of the bootstrap distribution adjusted for bias correction) for the partial correlation (\ref{eq:partialMantel}) is larger than 0. 

\subsection{The linear mixed effects model}\label{theLMM}
Another approach to model the dependence between dissimilarities involving the same individual is via introducing individual random effects into a regression between geographical and genetic dissimilarities.

\cite{clarke2002} proposed such a model. They were working with population-level genetic and geographical data. After centering the geographical distances to remove correlation between the intercept and slope estimates, they extended the linear regression between genetic and (log-transformed) geographical distances by introducing one random effect for each of the two populations on which the dissimilarity value was based. With the notation defined above and considering an individual-level analysis, it is possible to specify their model as
\begin{equation}
\label{clarke2002}
    d_y(\mathbf{z}_r, \mathbf{z}_c) = a + b\left(d_x(\mathbf{z}_r, \mathbf{z}_c) - \bar{d}_x \right) + \tau_r + \tau_c + \epsilon(\mathbf{z}_r, \mathbf{z}_c) \quad \text{with}\, n\geq r>c \,,
\end{equation}
where $a$ is a constant term, $\tau_i$ ($i=r$ or $i=c$) is a random effect representing the average deviation of $d_y$ values involving population $i$ from that expected from its $d_x$ distances to the other populations and $\tau_i$ and $\epsilon(\mathbf{z}_r, \mathbf{z}_c)$ are assumed to be independent with $\epsilon(\mathbf{z}_r, \mathbf{z}_c)$ i.i.d. normally distributed. This specification assumes that dependence between two dissimilarities involving the same population can be expressed by an additive random value. Technically this allows for dissimilarities smaller than zero, and does not take into account dependence that involves more than two pairs of populations, as exists for distances at least due to the triangle inequality. The model can therefore not be fully correct for a regression between distances, but given that all models are idealisations and simplifications, the model can still be suitable if it allows for inference with good performance characteristics. 

The authors fitted the model via restricted maximum likelihood (REML). It has gained popularity in the landscape genetics literature \citep{peterman2021}, being used to assess the effect of landscape variables on gene flow \citep{vanstrien2012} and for landscape model selection \citep{shirk2017}. It can be fitted using the \texttt{mlpe\_rga} function from the \texttt{ResistanceGA R} package \citep{resistanceGA}, based on the \texttt{lme4} package \citep{lme4}.

In order to apply model (\ref{clarke2002}) to species delimitation, a fixed effect associated with the grouping distance $\mathbf{D}_g$ needs to be incorporated:
\begin{equation}
\label{ourMLPE}
    d_y(\mathbf{z}_r, \mathbf{z}_c) = b_0 + b_1 d_x(\mathbf{z}_r, \mathbf{z}_c) + b_2 d_g(\mathbf{z}_r, \mathbf{z}_c) + \tau_r + \tau_c + \epsilon(\mathbf{z}_r, \mathbf{z}_c) \quad \text{with}\, n\geq r>c \,.
\end{equation}
$b_2$ here is an intercept update for between group genetic dissimilarities, and the null hypothesis $b_2=0$ corresponds to conspecificity, which is tested against the one-sided alternative $b_2>0$. In Figure \ref{fig:brassydd}, $b_2$ would be the amount by which the green between-groups dissimilarities are on average higher than the red and black within-group dissimilarities. This is similar to $H_{02}$ in Section \ref{sec:hh20}, assuming implicitly that there is no difference between the within-group regressions for group 1 and group 2. Even if there is such a difference, it can be seen as relevant to whether the dissimilarities between groups are systematically larger than a model defined on the aggregated within-group dissimilarities.

Note that different from the approaches in Sections \ref{sec:hh20} and \ref{thePMT}, (\ref{ourMLPE}) provides a \textit{generative} model for dissimilarities, but we will not use it as such because it does not use information about the underlying genetic dissimilarities, and also, as argued above, it cannot be fully correct for these. 

The test can be based on profile likelihood-based confidence intervals (CI) \citep{venzon1988, royston2007}. The null hypothesis is rejected if the lower boundary of the $(1-2\alpha)$ profile likelihood-based CI is larger than zero. 
These CIs are obtained in R via the \texttt{confint} function applied on the \texttt{mlpe\_rga} output. Profile-likelihood-based CIs are connected to likelihood ratio tests. Therefore, since inference involves a fixed effect, model (\ref{ourMLPE}) should not be fitted with REML \citep{west2022}.

A related approach was used by \cite{yang2004} for estimating and testing for isolation by distance. Instead of introducing random effects explicitly, several standard correlation patterns for the $\epsilon(\mathbf{z}_r, \mathbf{z}_c)$ as available in the SAS PROC MIXED \citep{SAS} were used to model the dependence in the dissimilarities. This can be expected to be inferior to (\ref{ourMLPE}), because it does not use the information which dissimilarities belong to the same individual.

\paragraph{A linear regression model ignoring dependence}
In our study we will also compare a straight linear regression model without the random effects that ignores the dependence between dissimilarities:
\begin{equation}
    \label{eq:IIDregr}
    d_y(\mathbf{z}_r, \mathbf{z}_c) = b^*_0 + b^*_1 d_x(\mathbf{z}_r, \mathbf{z}_c) + b^*_2 d_g(\mathbf{z}_r, \mathbf{z}_c) + \epsilon(\mathbf{z}_r, \mathbf{z}_c) \quad \text{with}\, n\geq r>c \,,
\end{equation}
assuming $\epsilon(\mathbf{z}_r, \mathbf{z}_c)$ i.i.d. normally distributed with zero mean. The null hypothesis is once more $b_2^*=0$, tested against $b_2^*>0$ with the standard regression t-test. A similar approach has been taken for distances in \cite{spriggs2019}. Note that if (\ref{eq:IIDregr}) held indeed, 
the null hypothesis $\rho(\mathbf{D}_y, \mathbf{D}_g | \mathbf{D}_x)=0$ of the partial Mantel test would be equivalent to $b_2^*=0$ \citep{legendre2000}.

\section{Simulations}\label{ch:simulations}

The simulators employed in this study simulate genetic data for individuals with geographical locations rather than dissimilarities, see Figure \ref{fig:geneticdisflow}. They are based on models for the evolutionary processes that lie behind the modification of the alleles in the loci sampled from the individuals' DNA. These algorithms simulate the life cycle of individuals that inhabit an artificial map and, generation by generation, exchange their genetic material through migration schemes that give rise to different IBD behaviours. The genetic make-up of the output individuals is the result of this complex set of factors, which will indirectly impact the relationship between genetic and geographical distances.

The choice of parameter values for complex simulations like these is a challenging task. While it is crucial to inform this choice via the best available knowledge from real data \citep{adrion2020}, retrieving relevant information for the specific landscape model of interest can be hard. As an indication, \cite{pope2015} found that almost one third of the publicly available datasets they surveyed could not be recreated, while $40\%$ did not report geographical information; on top of this, only a share of the reproducible datasets would contain co-dominant markers compatible with the computation of shared allele distances. 
We mostly focus on simple archetypical situations that allow for a good understanding of the key factors that influence the methods' performances. In Section \ref{simreal} we present a simulation that was designed based on two groups present in the real dataset presented in Section \ref{ch:brassy}.


All tests are run at level $\alpha=0.05$. In the following, the two simulators used in this comparative study and the corresponding results are described.
\subsection{Simulations based on the SLiM simulator}
As explained in its user manual \citep{SLiM_manual}, SLiM \citep{slim2023} is a forward-in-time simulation package for constructing genetically explicit individual-based evolutionary models. Its default settings include non overlapping generations, diploid individuals and offspring generated by recombination of parental chromosomes with the addition of new mutations. Within each species, an arbitrary number of subpopulations can be simulated, connected by any pattern of migration. Mutations at specific base positions in the genomes are explicitly modelled, also as nucleotide sequences. SLiM provides support for continuous space, either in one, two or three dimensions, and this can help simulate dispersal (mate choice with spatial kernel or nearest-neighbor search) and spatial competition. Importantly, SLiM allows the simulation of more than one species in a single SLiM model, opening the door to ecological interactions and coevolutionary dynamics. Virtually any feature of the simulated evolutionary scenario can be controlled via the integrated Eidos scripting language, which was created specifically for SLiM.

The short summary above hints at how its numerous simulation possibilities could be exploited to investigate the type I error rate and power of the methodologies explained in the previous sections. 
A two-group continuous space simulation, with individuals that compete for foraging areas (resulting in a more likely reproduction of  more isolated individuals), choose mates among their nearest neighbors and generate offspring in their surroundings, will lead to observations isolated by distance. Instead, absence of competition and less parent-dependent offspring positioning will lead to quasi-panmictic results, i.e., to the lack of association between genetic and geographical distances. The individuals from the two groups might inhabit the same geographical area or might be segregated into two disjoint areas. As regards conspecificity and distinctness, we used five different scenarios:
\begin{itemize}
    \item A simulation with one species and one subpopulation sampled with an artificial random split will return a naive conspecificity scenario, which can be used as baseline, referred to as \textit{random split (rs)}. 
    \item a simulation with only one species and two subpopulations, which descend from the same parent population and are able to exchange migrants, may yield a scenario consistent with the null hypothesis of conspecificity (referred to as \textit{conspecifity}, with two versions explained below), because the two groups are related by their history and their individuals are expected to display a similar genetic make-up;
    \item a simulation exploiting SLiM's multispecies engine (introduced in chapter 19 of the manual), with two distinct species that cannot interact, will generate a scenario consistent with the alternative hypothesis of distinctness (referred to as \textit{distinctness}, with two versions as explained below), since their genetic information is expected to be completely unrelated.
\end{itemize}
The details about SLiM's assumptions and key parameters for the simulations carried out in this study are reported below. The code can be found in the Supplementary Material. 

The simulation was initialized on a two-dimensional map and with an explicit nucleotide sequence 1000 base pairs long: this means that a total of 1000 diploid loci (technically, two genomes with 1000 positions) were simulated, where four different alleles can be found - corresponding to the four nucleobases: adenine, cytosine, guanine and thymine. The initial nucleotide sequence was random and the recombination rate was set to the default low value of $10^{-8}$ (loci were not independent). Mutations were handled according to the Jukes-Cantor mutational model \citep{jukes1969} by specifying a matrix containing the mutation rates from one nucleobase to the other: a unique rate applied to all transitions among nucleobases and in these simulations it was set to $0.0025$, a value that is larger than the default: too low values would lead to too few mutations and thus less genetic structure, given the timescale of the simulation. The map was always a square $200\times 200$ units wide, but in any case species were not allowed to get out of the central $100\times 100$ area. As far as mating is concerned, in all scenarios individuals would randomly pick a mate among their three closest neighbors, selected within a circular area of radius $3$.

On top of these shared parameter options, the following settings varied according to the scenario:
\begin{itemize}
    \item In the \textit{random split} scenario, throughout the simulation, all individuals inhabited the central $100\times 100$ area. In the \textit{conspecificity} scenario, the parent population inhabited the central square area, but, depending on the version, the two \say{children} subpopulations would continue to share the same wide area (\textit{overlapping conspecific (oc)}) or start to migrate to two disjoint areas of the map (\textit{separated conspecific (sc)}). By the time of the last simulated generation, the first subpopulation would inhabit the area between point $(50, 50)$ and point $(100, 100)$, whereas the second group would inhabit the area between point $(100, 100)$ and point $(150, 150)$, both included in the original wide central square area. Also in the \textit{distinctness} scenario there were two versions: in the \textit{overlapping distinct (od)} one, both species would inhabit the central $100\times 100$ area, whereas in the \textit{separated distinct (sd)} one, they would inhabit, since the very first generation, the area between point $(50, 50)$ and point $(100, 100)$ and the area between point $(100, 100)$ and point $(150, 150)$, respectively. 
    \item In all simulations, SLiM would output the data regarding the individuals only at the last generation. In the \textit{random split} scenario, one hundred generations were simulated. In the \textit{conspecificity} scenarios, the parent population would be simulated for the first $100$ generations and be removed afterwards; the two children subpopulations would originate from the parent one at generation $90$ and then be simulated up to generation $150$. In the \textit{distinctness} scenarios, both species would be simulated for $50$ generations. 
    \item The only subpopulation existing in the \textit{ random split} scenario and the parent population in the \textit{conspecificity} scenarios were made up of $400$ individuals. The children subpopulations in the \textit{conspecificity} scenarios and the two species in the \textit{distinctness} scenarios were made up of $200$ individuals each. Note that, in each SLiM cycle, individuals are born, mate and die, but the default behaviour of the software is to keep the number of simulated individuals steady throughout the generations. 
    \item In order to generate diverse data richness situations, regardless of the scenario, the number of loci available for the computation of the shared allele distance was either $4$, $40$ or $89$ (randomly selected) and the total number of individuals sampled was either $12$, $30$ or $90$. The whole simulation study was carried out either with equally sized groups (e.g., $6$ versus $6$ individuals) or with one group being twice as large as the other (e.g., $4$ versus $8$). In the \textit{random split} scenario there was just one big group from which individuals were drawn, and these individuals were then randomly assigned to the two groups used to test conspecificity. Given the membership, the drawing of individuals was random, except for the \textit{separated conspecific} scenario, when it was constrained to those individuals inhabiting the subpopulation-specific geographical area of the map: indeed, given the migration process involved in that scenario, it could well happen that some individuals at the last generation were still positioned in the area specific to the other group, typically because one of their parents belonged there. 
    \item As far as spatial competition is concerned, it was modeled through the effect that interactions between individuals had on their probability to reproduce. Each individual experienced an interaction strength that was the sum of all its interactions with individuals in the neighbourhood. In particular, a Gaussian kernel was used to translate the distance between two individuals in the strength of the interaction between them: when trying to enforce a strong IBD behaviour in the individuals, this Gaussian distribution would have mean $2.5$ and standard deviation $5$ and the interactions with individuals out of the circular area of radius $15$ would be set to zero; when trying to mimic no IBD species, the distribution would have mean $0.5$ and standard deviation $1$ and the circular area with positive-valued interactions would have radius $3$. With the first parameter settings, given a certain Euclidean distance between two individuals, the strength of the interaction would be assigned a larger value: the stronger the total interaction felt by an individual, the lower its probability to reproduce, leading to the formation of isolated subgroups and hence to restricted gene flow. With narrower Gaussian kernels, instead, the total interaction strength on each individual would tend to be smaller and thus there would be less incentive to dispersal, resulting in a more panmictic-like behaviour. 
    \item In the \textit{conspecificity} scenarios, the two children subpopulations were allowed to exchange migrants. A migration rate of $20\%$ means that when creating the offspring for, say, the first subpopulation, $20\%$ of the parents (with some stochastic variability) were picked from individuals belonging to the second subpopulation.
In the \textit{overlapping conspecific} scenario, the two subpopulations would exchange parents at a rate randomly oscillating between $40$ and $50\%$ till the last generation.
    In the \textit{separated conspecific} scenario, the migration rate would start off at $20\%$ and then linearly decrease to reach zero in the last generation, at a pace that is consistent with the progressive separation of the geographical areas.
    \item As far as offspring generation is concerned, it occurred at every simulation cycle after the choice of the two mating parents: its position was shifted from that of the first parent according to a draw from a zero-mean Gaussian kernel with standard deviation $1$ in case of strong IBD behaviour and $9$ in case of quasi-panmictic behaviour. Thus, with strong spatial competition, the emerging isolated groups would tend to be preserved because offspring were more likely to emerge in a narrow neighbourhood of their parents. In the \textit{separated conspecific} scenario, the location parameter of the Gaussian distribution involved in this process was modified according to the group: for the first group, that would end up in the square area between point $(50, 50)$ and point $(100, 100)$, the parameter was set to $-0.5$, whereas it was equal to $0.5$ for the second group. Also in this respect, this is consistent with the gradual process of separation that affected the groups since the $100^{\text{th}}$ generation. 
\end{itemize}
In the Supplementary Material, example distance-distance plots similar to Figure \ref{fig:brassydd} from the scenarios are shown, with equal or unequal group sizes, both for quasi-panmictic groups and for isolated by distance groups.

\paragraph{Results from SLiM}
Five scenarios were simulated with SLiM: \textit{rs} (one group, random split), \textit{oc} (two groups, same parent population, same inhabited area), {sc} (two groups, same parent population, disjoint inhabited areas), \textit{sd} (two species living in disjoint areas) and \textit{od} (two species inhabiting the same area). Across the scenarios sorted in this way, the rejection rate from species delimitation methods is expected to be non-decreasing, since we transition from a clear conspecificity setup (\textit{rs}) to a clear distinctness setup (\textit{od}); if species inhabit the same area, both conspecificity  and distinctness can be more easily diagnosed, so that we expect a lower rejection rate for \textit{oc} than for \textit{sc}, but a higher one for \textit{od} than for \textit{sd}. On top of these scenarios, other varying parameters (all shared by both groups) were the IBD behaviour and the number of loci available for the computation of $\mathbf{D}_y$ (out of the $1000$ loci simulated). In half of the cases the two groups were equally sized and in the other half $n_2 = 2 n_1$. The combination of all these factors generated 36 scenarios and in each of them the techniques described in Section \ref{ch:methods} were applied both with untransformed and log-transformed geographical distances. One hundred replicates of each of these combinations were generated and the number of rejections was recorded for all the methods. This information is visualized in power plots, one for equal and one for unequal group sizes. In these plots, HH20 denotes the protocol by \cite{hh20}; PMantel denotes PMT with permutations; MJack denotes PMT with jackknife; BC denotes PMT with bias-corrected bootstrap; MXD denotes the mixed effects model (\ref{ourMLPE}); LM is the multiple regression ignoring dependence.

\begin{figure}[tb] 
    \centering
    \includegraphics[width=1\textwidth]{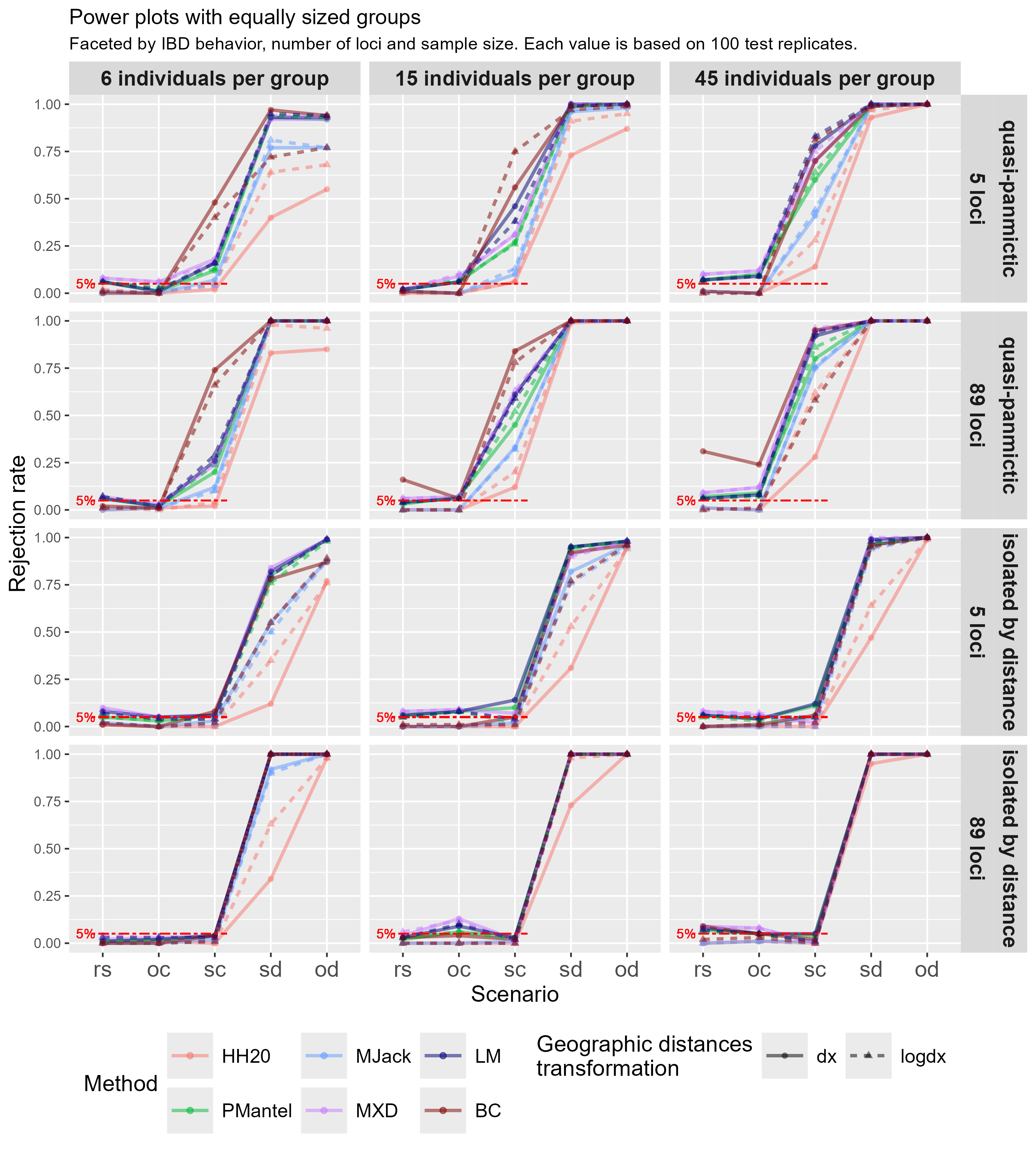}
    \captionsetup{width=.9\linewidth}
    \caption{Power plot based on \textbf{SLiM} data simulated \textbf{with equally sized groups}. Panels are done by number of individuals per group (rows) and a combination of IBD behavior and number of available loci (columns). Each rejection rate is based on 100 simulations with the same parameter settings. Circles and solid lines refer to analyses with untransformed geographical distances, whereas triangles and dashed lines refer to analyses with log-transformed ones. The horizontal dashed red line is superimposed to help assess type I error rates. }
    \label{fig:SLiMpower_equal}
\end{figure}
In Figure \ref{fig:SLiMpower_equal}, rejection rates from the scenarios with equally sized groups are reported (methods' abbreviations are explained in the caption). Results with $40$ available loci are only shown in the Supplementary Material, same regarding Section \ref{GSpace_section}. The expected non-decreasing trend in the rejection rates was confirmed, with some minor exceptions between the \textit{rs} and the \textit{oc} (same parent, same area) scenario. In the \textit{rs} scenario, all methods (surprisingly, model (\ref{eq:IIDregr}), too) displayed a type I error rate close to the significance level $5\%$, although the bias-corrected bootstrap with untransformed $\mathbf{D}_x$ rejected the null hypothesis of conspecificity too often in some setups. In general, the jackknife-based methods (HH20, MJack) showed type I error rates very close to zero, whereas all other methods had them slightly above the significance threshold. In the second scenario, especially with large samples, these methods showed rejection rates close to $10\%$.

In \textit{sc} (same parent, disjoint areas) with quasi-panmictic groups, all rejection rates registered a strong increase: especially with many individuals and loci, all methods seemed to suggest that the two groups should be considered distinct, despite having originated from the same parent population. This was probably due to the combination of geographical separation and absence of IBD behaviour: the genetic structure that was formed because of the decreasing migration rate could not be explained by geographical distance as individuals in the same group tended to be quasi-panmictic. Indeed, when species were isolated by distance and with sufficient genetic information ($89$ available loci), the rejection rates in the \textit{sc} scenario all fell below the significance level. Recalling the remark in the Introduction about different levels of biological differentiation, it can be controversial whether the groups generated with this particular SLiM script should be considered conspecific. Most methods concluded they are not, which is important information for biologists using these tests.

Under the multispecies setups, all methods displayed a rejection rate close to 1. The jackknife-based methods, though, tended to display lower power than the other methods, particularly with small sample sizes. This trend was common to all scenarios: PMantel, MXD, LM and BC always showed rejection rates larger than HH20 and MJack. In this respect, it is worth noting that MJack, representing a compromise between HH20 and PMantel, displayed satisfactory type I error rate and larger power than HH20 in all setups. Now, if the rejection of the null hypothesis in \textit{rs} is seen as a Bernoulli random variable with success probability equal to $0.05$, 10 rejections out of 100 would represent a significant result at 5\% level. PMantel did not consistently display such significant figures in the simulations, but in an ad hoc simulation under the null hypothesis, with $15$ individuals per group and $40$ loci (not shown here), this test rejected 70 times out of 1000 repetitions (one-sided p-value $= 0.0023$ against the null hypothesis that the rejection probability is 0.05).
By replacing the permutation-based significance assessment with a jackknife-based one, MJack achieved more power than HH20 while keeping its same low type I error rate. This might support the idea that permutations introduce a distortion in the distribution of geographical distances.

\begin{figure}[tb] 
    \centering
    \includegraphics[width=1\textwidth]{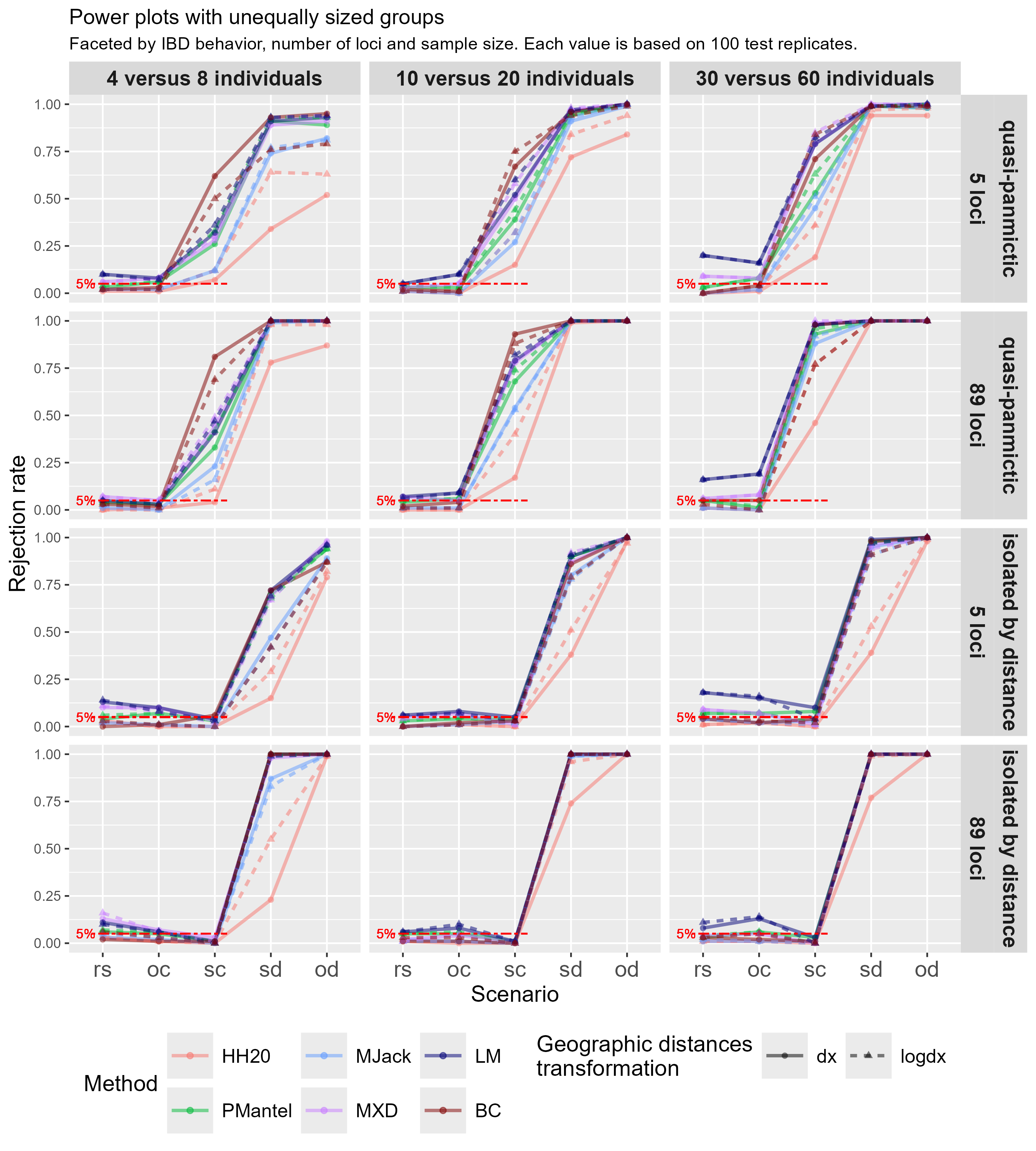}
    \captionsetup{width=.9\linewidth}
    \caption{Power plot based on \textbf{SLiM} data simulated \textbf{with unequally sized groups}. See the caption to Figure \ref{fig:SLiMpower_equal} for further description.}
    \label{fig:SLiMpower_unequal}
\end{figure}
Also in Figure \ref{fig:SLiMpower_unequal}, with unequal group sizes, the rejection rates were mostly non-decreasing when going from \textit{rs} to \textit{od}. The most important difference regards the type I error rate of LM: when one group was twice as large as the other, the rejection rate of LM in the \textit{rs} scenario often was above the significance level, sometimes strongly so, especially with the largest sample sizes.

Regardless of the other parameter options, the transformation of the geographical distances did not seem to have a relevant impact on the methods' performance. The only exception is HH20, whose power increased with log-transformed geographical dissimilarities. 

\subsection{Simulations based on the GSpace simulator}\label{GSpace_section}

GSpace is a coalescent-based simulator designed to model genetic variation in spatially structured populations under isolation by distance. It reconstructs the genealogical history of sampled individuals while accounting for recombination and migration across a lattice-based landscape. The lattice structure allows for flexible spatial modeling, where each vertex on the lattice can represent a local population or the position of individual organisms in a continuous habitat approximation.

The simulation proceeds backward in time: starting from the sampled individuals, their genealogies are traced until the most recent common ancestor (MRCA) is reached. At each generation, migration is modeled using two-dimensional dispersal distributions (e.g., uniform, geometric, or Zeta), determining how individuals move between locations. Once the MRCA is identified, mutations are introduced forward in time along the branches of the coalescence tree.
Further details can be found in \cite{ibdsim2009}.

As far as the specific settings for the simulations carried out in this study are concerned, a $200\times200$ vertices map was created, with diploid individuals always inhabiting the central $60\times60$ area. A first group would inhabit the area between vertex $(70, 70)$ and vertex $(90, 90)$, whereas a second group would inhabit the square between vertices $(110, 110)$ and $(130, 130)$. Their coordinates were drawn uniformly within the allowed range. A sequence of di-allelic loci was simulated, with both the mutation rate per generation per locus and the recombination rate per generation between loci equal to $0.005$ (ten times the default) for all simulated loci. The so-called K-allele mutation model was used, according to which the initial allelic state is changed into one of the other possible states - in this case the only other allele allowed in the di-allelic setting. In order to mimic a continuous habitat, in each vertex there could be up to one individual.

In addition to this, the following parameters varied according to the scenario:

\begin{itemize}
    \item In order to obtain a baseline scenario where the conspecificity hypothesis was trivially true, the two groups were simulated within the same software execution, so that the algorithm would reconstruct a unique genealogy common to all individuals (referred to as \textit{1} in the results plots). In all other cases \textit{2s}, \textit{2o}, \textit{2n}, see below), the two geographically separated groups were generated during two distinct software executions and collected in the same dataset. Note that the baseline scenario in SLiM had the two groups inhabiting the same area, unlike what is done here with GSpace. 
    \item The two groups would obviously share the same allele pools (the same two allele options for all loci) when generated within the same software execution, whereas they could share both alleles (referred to as \textit{2s}, ``s'' for ``same allele pool''), one allele (\textit{2o}) or no allele (\textit{2n}) otherwise. Of course, when no allele was shared, the genetic dissimilarities between individuals from different groups would always take value one. 
    \item IBD behaviour was controlled via the choice of the univariate dispersal distribution: GSpace assumes that dispersal occurs independently in each dimension. In order to yield panmictic groups, a uniform dispersal was used, according to which the probability of moving $t$ steps in one direction is $\frac{m}{d_{max}}$, where $m$ is the total migration rate, equal to $0.5$ and $d_{max}$ is the maximum distance reachable in any migration event, set to $200$ (the largest possible value) in all scenarios. As regards IBD species, a Zeta (or truncated discrete Pareto) dispersal distribution was used, assigning value $\frac{m}{2 {|t|}^{\kappa}}$ to the probability to move $t$ steps in one direction, with $\kappa=5$ being the shape parameter.
    \item The total number of simulated individuals was either $12$, $30$ or $90$, whereas the genetic sequence was either $4$, $40$ or $100$ loci long. As with SLiM, all scenarios were investigated both with equally sized groups and with one group twice as large as the other. 
\end{itemize}
The script on which these simulations were based and distance-distance plots from all simulated scenarios can be found in the Supplementary Material.

\paragraph{Results from GSpace}
The combination of the parameter settings illustrated above led to a total of four scenarios: the simulation involving a unique software execution represented a conspecificity setup, whereas the situation where the allele pools of the two groups shared no allele constituted a distinctness one. The other two setups were included as intermediate situations, with the two groups not sharing ancestors while still showing similarities in their genetic information. Recall that in all these scenarios the geographical separation was always the same: the two groups inhabited two disjoint areas of the map. The four scenarios can be sorted as follows: one execution (\textit{1}), two executions and same alleles (\textit{2s}), two executions and one allele in common (\textit{2o}), two executions and no allele in common (\textit{2n}). In this order, the rejection rate is expected to be non-decreasing.

On top of these scenarios, other varying parameters were the number of individuals per group, the number of loci available for the analysis and the IBD behaviour (absent with Uniform dispersal distribution and strong with Zeta dispersal distribution). Each of these parameter (and scenario) combinations was simulated one hundred times and the number of rejections was recorded. 
\begin{figure}[tb] 
    \centering
    \includegraphics[width=1\textwidth]{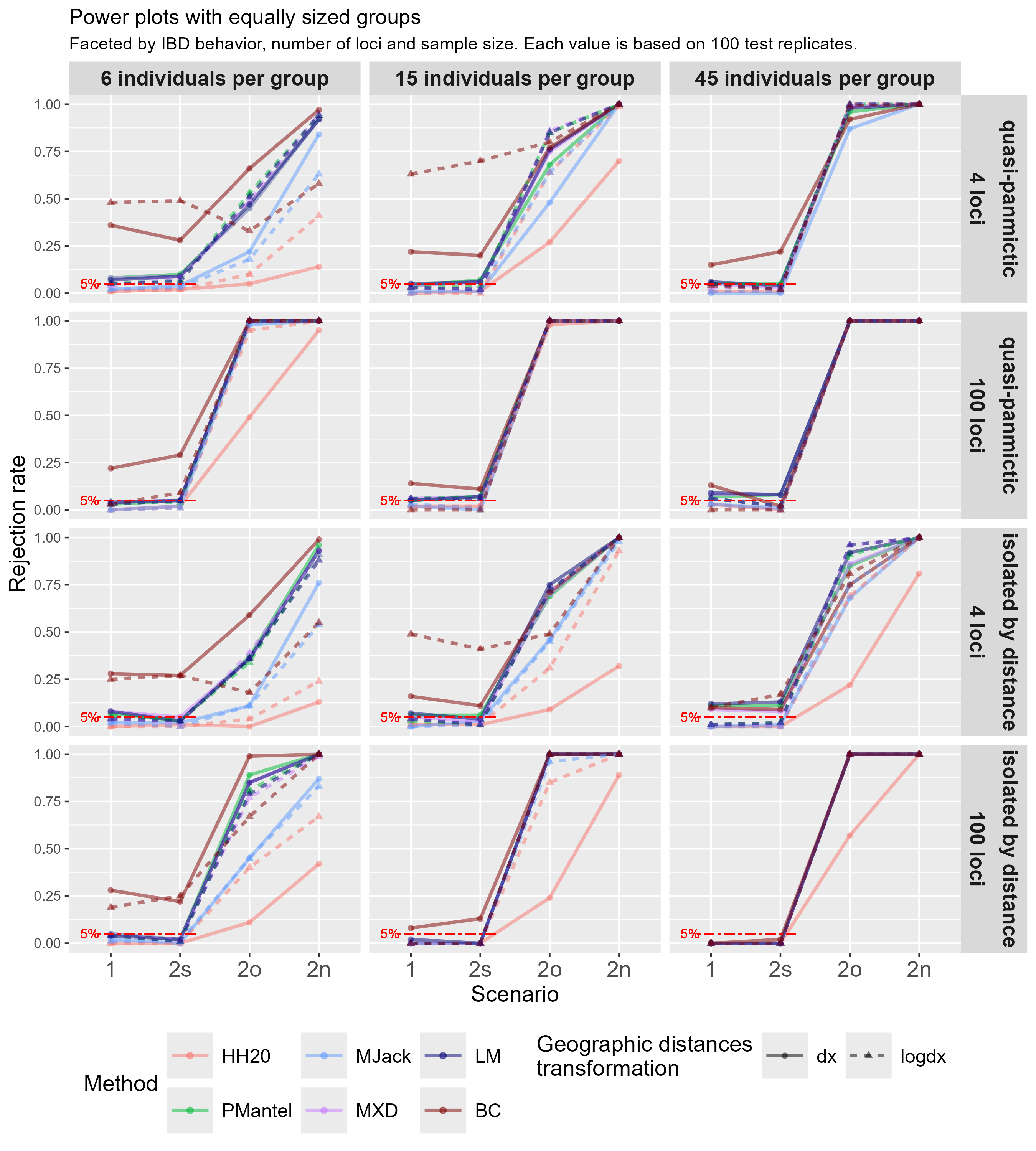}
    \captionsetup{width=.9\linewidth}
    \caption{Power plot based on \textbf{GSpace} data simulated \textbf{with equally sized groups}. Panels by combination of IBD behaviour and number of loci (columns) and number of individuals per group (rows). See the caption to Figure \ref{fig:SLiMpower_equal} for further description.}
    \label{fig:GSpacepower_equal}
\end{figure}

In Figure \ref{fig:GSpacepower_equal} the results related to situations in which the two groups are equally sized are reported. It is apparent how the ranking of the methods in terms of performance was similar to that observed in SLiM, apart from the odd behaviour of BC, which displayed very large type I error rates especially with fewer loci and fewer individuals. This is due to resampling with replacement, which led to the emergence of too many zero dissimilarities in the bootstrap samples. A closer look at the related distance-distance plots (not shown here) suggests that these zero dissimilarities were outlying with respect to the bulk of the data, leading to a biased distribution of bootstrap replicates of the partial correlation coefficient. Indeed, accounting for geographical separation, in those situations there appeared to be a large positive association between genetic and grouping distance, as zero distances could only be found when the grouping indicator was equal to zero. Consequently, bootstrap replicates had an upward bias that often caused a rejection of the null hypothesis. This phenomenon was not evident with SLiM data because its scenario \textit{rs} did not involve geographically separated groups, unlike GSpace. These findings represent empirical evidence that partial Mantel testing with bootstrap can generate unreliable results.

The jackknife-based methods showed lower power with respect to other methods, and once again the logarithmic transformation of $\mathbf{D}_x$ helped HH20 to be more powerful. In scenario \textit{2o}, given the number of loci and the total sample size, the rejection rates were lower with IBD species than with quasi-panmictic ones. Indeed, with species isolated by distance, geographical separation was sometimes enough to explain the genetic discrepancies between the two groups, and jackknife-based methods were more sensitive to this data feature than other methods.
\begin{figure}[tb]
    \centering
    \includegraphics[width=1\textwidth]{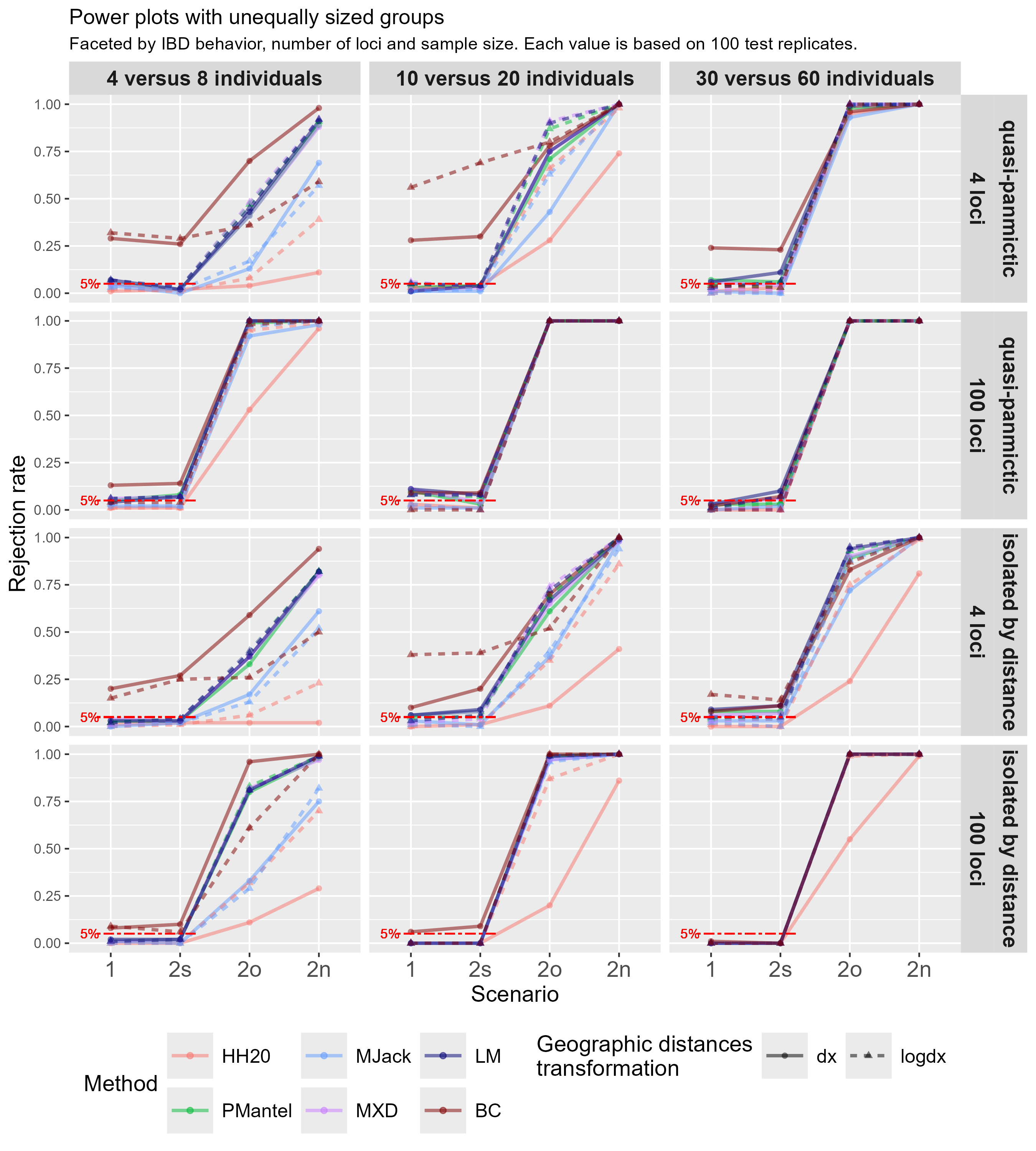}
    \captionsetup{width=.9\linewidth}
    \caption{Power plot based on \textbf{GSpace} data simulated \textbf{with unequal group sizes}. Panels by combination of IBD behaviour and number of loci (columns) and number of individuals per group (rows). See the caption to Figure \ref{fig:SLiMpower_equal} for further description.}
    \label{fig:GSpacepower_unequal}
\end{figure}

Figure \ref{fig:GSpacepower_unequal} shows that, when the two groups had unequal sizes, the rejection rates were similar to those with equally sized groups. A slight decrease of some rejection rates across all scenarios can be spotted in the column with IBD species and four loci.

\subsection{A simulation inspired by real data}\label{simreal}
In order to incorporate a simulation using realistic parameters backed up by real data, we considered the Cassioides and Tyndarus subgroups of the brassy ringlets data analyzed and discussed in more detail in Section \ref{ch:brassy} and shown in the last plot in Figure \ref{fig:brassydd}.

For this task, GSpace was preferred to the SLiM simulator because the latter does not allow to fix the geographic positions of the individuals in advance and involves many more parameters to be set.

Geographical coordinates of the individuals (available in the dataset) were converted to integers and mapped on the vertices of the GSpace lattice. For each of these two subgroup of brassy ringlets, 32 parameter combinations were compared (narrowed down from earlier experiments with 300 combinations) in order to find the parameter settings that best mimicked the original trend in the dissimilarities. For details on how the best parameters were found, consider the Supplementary Material, which also has the GSpace script. As a result, for both subgroups, a Zeta distribution with $\kappa=1$, eight unique allelic states, the KAM mutation model, a recombination rate of $5^{-7}$, and the mutation rate of $10^{-6}$ were used. 

As shown in Section \ref{ch:brassy}, the evidence is very strong that the Cassioides and Tyndarus subgroups constitute two different species. 
In a first simulation, individuals from two groups were simulated using the parameter settings above and based on the original geographical positions, but during two distinct software executions. 100 such datasets were simulated. We assume that in this situation the subgroups are indeed different species, so that a rejection of the null hypothesis is correct. Estimating the rejection probability therefore amounts to measuring the power of the tests.

In a second simulation, we simulated a single group of 36 individuals (as many as in both subgroups combined) with the above parameters. We then splitted individuals into two subgroups according to their original membership (Cassioides vs Tyndarus) in the real data. This simulates a single species, so that a rejection of the null hypothesis constitutes a type I error.

\begin{table}[tb]
    \captionsetup{width=0.9\linewidth}
    \caption{Empirical rejection rates of the methodologies discussed in Section \ref{ch:methods} for the first (power) and second (type I error) simulation described in Section \ref{simreal} based on 100 datasets each.}
    \centering
    \begin{tabular}{l|llllll}
    Method & HH20 & PMantel & MJack & MXD & LM & BC \\
    \hline
    Power & $0.84$ & $0.96$ & $0.94$ & $1$ & $0.98$ & $0.94$ \\
    Type I error & $0.03$ & $0.11$ & $0.03$ & $0.18$ & $0.22$ & $0.04$
    \end{tabular}
    \label{tyncas_levelpower}
\end{table}

The results of both simulations are given in Table \ref{tyncas_levelpower}. Largely in line with the earlier simulation results, HH20, MJack, and BC achieved a type I error probability below 0.05, whereas the other three methods rejected clearly too often in the type I error simulation, which makes their good power useless; MXD was worse here than in the other simulations. Once more, MJack achieved a clearly higher power than HH20.



\subsection{Discussion}
The individual-based spatially-explicit simulations carried out via SLiM and GSpace allowed to study the type I error rate and power of the techniques described in the \nameref{ch:methods} section. A consistent ranking in the overall performance of the methodologies could be observed, with jackknife-based methods being more conservative and less powerful and the other techniques being more powerful, but at the cost of occasional type I error rates above the significance level.

By combining jackknife significance assessment and the usage of partial correlation coefficients, MJack managed to achieve better power than HH20, while showing type I error rates consistently below the significance level, unlike PMantel. The bias-corrected bootstrap-based PMT showed a too large type I error rate in the most challenging data setups. No method apart from HH20 clearly benefited from the logarithmic transformation of the geographical distances. Despite the trend in the dissimilarities often being convex, the assumption of linearity did not seem to have a severe impact on performance.

Despite wrongly assuming that all dissimilarities in the dataset are independent, OLS estimation in model (\ref{eq:IIDregr}) displayed good type I error rate and power in several situations. In particular, it often performed better than the random effects model MXD, avoiding certain increased type I error probabilities of the latter. However, with SLiM data, the type I error rate of LM was consistently above the significance level when unequally sized groups were being compared. This is in line with the expectation that ignoring the dependence between distances should lead to an underestimation of variability.

The performance of HH20 was summarized by a single rejection rate in the power plots, but recall that this method is hierarchical. In our study, a rejection corresponded to cases when either $H_{02}$ or $H_{03}$ was rejected, and thus, non-rejections included also the inconclusive results that can arise when testing $H_{03}$. Although in all simulations both groups always had the same IBD behaviour, some rejections of $H_{01}$ occurred, so that $H_{03}$ rather than $H_{02}$ was tested. The possibility to take into consideration unequal IBD behaviour is unique to HH20, but here may have led to a degradation of its performance. More precisely, lower rejection rates may have been recorded here in spite of the fact that, in some unclear cases, a practitioner could have concluded that two distinct species were being compared.

\section{Analysis of brassy ringlets data}\label{ch:brassy}

\cite{gratton2016} discussed the biological delimitation of a taxon of butterflies (brassy ringlets; \textit{Erebia tyndarus} complex, Lepidoptera) endemic to Southern Europe, the Altai Republic and the Rocky Mountains. They studied the morphological, genetic and geographic information of $45$ individuals netted during the summer of 2012 across the Italian Appennines, the Alps and the Pyrenees. Four subgroups of this clade were represented in the sample, namely \textit{E. Tyndarus}, \textit{E. Nivalis}, \textit{E. Calcaria} and \textit{E. Cassioides}, with the latter possibly divisible in three populations according to the area of collection. After selecting a subset of $389$ diploid loci, they applied k-means clustering on the principal components obtained from the genetic data, Bayesian model-based clustering using the \texttt{STRUCTURE} software \citep{pritchard2000} and coalescent-based Bayes factor delimitation \citep{leache2014}, integrating their results by examining the isolation by distance behaviour and morphological differentiation of the individuals in each putative cluster. The study in \cite{gratton2016} did not only back up the distinction between the four groups mentioned above from a genetic point of view, but also supported further differentiation within the \textit{Cassioides} group among the Eastern and Orobian Alps population, the Southern and Central Appennines population and the population inhabiting Northern Appennines, Pyrenees and Western Alps. 

\cite{hh20} applied their testing protocol to these data in order to replicate and deepen the IBD investigation carried out by \cite{gratton2016}. With the exclusion of the \textit{Calcaria} group, which could not be examined due to the small sample size (only $3$ specimens), the distinction among the groups was confirmed. The classification within the \textit{Cassioides} group, instead, was slightly amended: HH20 suggested that there was no evidence of distinctness between the Southern and Central Appennines population and the population inhabiting Northern Appennines, Pyrenees and Western Alps, whereas the genetic dissimilarity between these populations taken together and the Eastern and Orobian Alps population was too large to be explained by isolation by distance.
\begin{table}[tb]
\footnotesize
    \caption{Results from all methods compared here on the brassy ringlets data. For the three tests in HH20, p-values are reported (two of them are given for $H_{03}$ when discordant); p-values are reported for PMantel, MJack and LM, too; for MXD and BC, confidence intervals are reported for the $b_2^*$ regression coefficient and the partial correlation coefficient, respectively: if their lower boundary is larger than zero, the null hypothesis is rejected.}
    \begin{tabular}{p{0.2\linewidth}|cccccccc}
    \hfill Groups compared & $H_{01}$ & $H_{02}$ & $H_{03}$ & PMantel & MJack & MXD & LM & BC \\[3pt]
    \hline \\
    \textit{E. Tyndarus} \textbf{vs} \textit{E. Nivalis}& $0.074$ & $<10^{-5}$ & n.a. & $0.001$ & $<10^{-29}$ & $(0.296, 0.306)$ & $<10^{-113}$ & $(0.983, 0.995)$ \\[3pt]
    \textit{E. Nivalis} \textbf{vs} \textit{E. Cassioides}& $0.094$ & $<10^{-4}$ & n.a. & $0.001$ & $<10^{-55}$ & $(0.378, 0.389)$ & $\sim 0$ & $(0.972, 0.988)$ \\[3pt]
    \textit{E. Tyndarus} \textbf{vs} \textit{E. Cassioides}& $<10^{-9}$ & n.a. & $\text{both} <10^{-24}$ & $0.001$ & $<10^{-67}$ & $(0.345, 0.352)$ & $\sim 0$ & $(0.977, 0.986)$\\[3pt]
    \textit{E. Cassioides}: W\_Alps + Pyrenees + N\_Apennines \textbf{vs} Orobian + E\_Alps & $0.487$ & $0.004$ & n.a. & $0.001$ & $<10^{-5}$ & $(0.025, 0.036)$ & $<10^{-17}$ & $(0.493, 0.745)$ \\[3pt]
    \textit{E. Cassioides}: W\_Alps + Pyrenees + N\_Apennines \textbf{vs} Central + S\_Apennines & $<10^{-4}$ & n.a. & $0.098; 0.004$ & $0.002$ & $0.015$ & $(0.030, 0.045)$ & $<10^{-5}$ & $(-0.148, 0.552)$ \\[3pt]
    \textit{E. Cassioides}: Central + S\_Apennines \textbf{vs} Orobian + E\_Alps & $0.144$ & $0.009$ & n.a. & $0.001$ & $<10^{-4}$ & $(0.041, 0.066)$ & $<10^{-15}$ & $( 0.477, 0.484)$ \\[3pt]
    \textit{E. Cassioides}: W\_Alps + Pyrenees + ALL Apennines \textbf{vs} Orobian + E\_Alps & $1$ & $<10^{-4}$ & n.a. & $0.001$ & $<10^{-8}$ & $(0.020, 0.029)$ & $<10^{-25}$ & $(0.329, 0.618)$ \\[3pt]
    \end{tabular}
    \label{tab:grattontable}
\end{table}

In Table \ref{tab:grattontable}, the results from all methods involved in this study are reported. See Figure \ref{fig:brassydd} for the dissimilarity data on which three of these comparisons are based.
The only non-rejections of the null hypothesis of conspecificity occurred for \textit{E. Cassioides}: W\_Alps + Pyrenees + N\_Apennines \textbf{vs} Central + S\_Apennines (second panel in Figure \ref{fig:brassydd}) by the bias-corrected bootstrap-based Partial Mantel test (the CI contains $0$) and the larger one of the two group-wise p-values for  $H_{03}$ by HH20. 

In all other cases, all methods agreed upon the distinctness results, confirming the conclusions shared by \cite{gratton2016} and \cite{hh20} - including also the distinction between the Western and Appennines populations versus the Eastern and Orobian populations (last row in Table \ref{tab:grattontable}). According to the evidence collected in this study, not only the \textit{E. Tyndarus}, \textit{E. Nivalis} and \textit{E. Cassioides} groups should be considered distinct: also the three subgroups identified within the \textit{E. Cassioides} group, namely the Eastern and Orobian Alps population, the Southern and Central Appennines population and the population inhabiting Northern Appennines, Pyrenees and Western Alps, display a genetic structure that cannot be explained by their geographic separation.

\section{Conclusions}\label{ch:conclusion}
We investigated methods that model the relationship between genetic and geographical dissimilarities in biological species in order to perform species delimitation. These techniques check whether the genetic structure existing between two putative species is compatible with the way genetic dissimilarities within each group increase with the geographical separation of the individuals. The type I error rate and power of these methods were compared by means of individual-based simulations carried out with the simulators GSpace and SLiM. Results showed that the method of \cite{hh20} (HH20) has a very conservative type I error rate and lower power than Partial Mantel tests (PMTs) as applied by \cite{medrano2014}, which in turn had type I error rates slightly above the significance level in some setups. Testing PMTs with jackknife instead of permutations fixed this behaviour, ensuring more power than HH20 while keeping the type I error rate still close to zero. This method can therefore be seen as the best in the simulated setups. Testing PMTs with bias-corrected bootstrap confidence intervals, instead, often led to inflated type I error rates.
The linear mixed effects model displayed a performance similar to PMT. A linear regression without random effects (LM), i.e., wrongly assuming independence among the dissimilarities, showed inflated type I error rates only with unequally sized groups simulated with SLiM, and performed surprisingly well otherwise.

The ranking in the overall performance of the methods was consistent over both simulators, and the log-transformation of the geographical dissimilarities did not seem to have a considerable impact on methods other than HH20, where it improved matters. The impact of transformations will in general depend on the used dissimilarities.

Due to the extremely large amount of possibilities defined by parameter choices of the simulators, many more potentially interesting situations could be simulated, and, as is always the case with such simulations, generalizability of results cannot be guaranteed beyond the simulated scenarios. Some scenarios worth exploring could involve comparisons between groups with different IBD behaviour, size and separation of the inhabited areas, but also simulations with independent loci, different settings for the time scale and migration rates, etc. The number of generations can be an important
parameter in the distinctness scenarios simulated with SLiM, but it is hard to investigate systematically. A too large number of generations can mean that shared allele distances between co-specific individuals might become so large that they could be indistinguishable from distances between species, and the effect of changing this may also interact with many other parameter choices.
Moreover, the investigation could be extended to population-based simulations, using genetic dissimilarity measures like $F_{st}$ \citep{weir1984} or the chord distance \citep{cavallisforza1967}. The individual-based methods compared in this study often assume independence between observations in the same group, something that is not biologically grounded and is indeed avoided in population-based studies. Independence between individuals is actually questionable even for the data simulated from GSpace and SLiM, where models account for contact between individuals within species. We do however think that it is realistic that statistical methods assuming independence are applied to real data that are not in fact truly independent, and such a situation is actually simulated.

In addition, another simulation-based investigation could be set up in order to compare the performance of these methods in scenarios where there is no putative grouping known in advance. To this end, an automated unsupervised routine for species delimitation may be conceived that uses the tests investigated here to decide whether groups that come out of a cluster analysis produced by methods such as \texttt{STRUCTURE} \citep{pritchard2000} should be merged. The \texttt{conStruct} approach \citep{bradburd2018} does something similar, but is only applicable when the number of loci is much larger than $n$, whereas, as shown in this study, most methods investigated here worked fairly good even with very small $n$ and $P$.

Applying the methods treated here to other problems of regression between dissimilarities such as relating similarity between languages or dialects to geographical distance \citep{bella21} may also be of interest.

Regarding the simulation in Section \ref{simreal}, despite using the best parameters we could find, there were still systematic differences between simulated and real data, so there may be further scope to improve the realism of the simulations. 

\addcontentsline{toc}{section}{References}

\printbibliography

@article{adriaensen2003,
title = {The application of ‘least-cost’ modelling as a functional landscape model},
journal = {Landscape and Urban Planning},
volume = {64},
number = {4},
pages = {233-247},
year = {2003},
issn = {0169-2046},
doi = {https://doi.org/10.1016/S0169-2046(02)00242-6},
url = {https://www.sciencedirect.com/science/article/pii/S0169204602002426},
author = {Adriaensen, F. and Chardon, J.P. and {De Blust}, G. and Swinnen, E. and Villalba, S. and Gulinck, H. and Matthysen, E.},
}

@article {adrion2020,
article_type = {journal},
title = {A community-maintained standard library of population genetic models},
author = {Adrion, J. R. and Cole, C. B. and Dukler, N. and Galloway, J. G. and Gladstein, A. L. and Gower, G. and Kyriazis, C. C. and Ragsdale, A. P. and Tsambos, G. and Baumdicker, F. and Carlson, J. and Cartwright, R. A. and Durvasula, A. and Gronau, I. and Kim, B. Y. and McKenzie, P. and Messer, P. W. and Noskova, E. and Ortega-Del Vecchyo, D. and Racimo, F. and Struck, T. J. and Gravel, S. and Gutenkunst, R. N. and Lohmueller, K. E. and Ralph, P. L. and Schrider, D. R. and Siepel, A. and Kelleher, J. and Kern, A. D.
},
editor = {Coop, G. and Wittkopp, P. J. and Novembre, J. and Sethuraman, A. and Mathieson, S.},
volume = 9,
year = 2020,
pub_date = {2020-06-23},
pages = {e54967},
citation = {eLife 2020;9:e54967},
doi = {10.7554/eLife.54967},
url = {https://doi.org/10.7554/eLife.54967},
journal = {eLife},
issn = {2050-084X},
publisher = {eLife Sciences Publications, Ltd},
}

@inbook{balkenholCH1,
author = {Balkenhol, N. and Cushman, S. A. and Storfer, A. and Waits, L. P.},
publisher = {John Wiley \& Sons, Ltd},
isbn = {9781118525258},
title = {Introduction to landscape genetics – concepts, methods, applications},
booktitle = {Landscape Genetics},
chapter = {1},
pages = {1-8},
doi = {https://doi.org/10.1002/9781118525258.ch01},
url = {https://onlinelibrary.wiley.com/doi/abs/10.1002/9781118525258.ch01},
eprint = {https://onlinelibrary.wiley.com/doi/pdf/10.1002/9781118525258.ch01},
year = {2015},
}

@inbook{balkenholCH3,
author = {Waits, L. P. and Storfer, A.},
publisher = {John Wiley \& Sons, Ltd},
isbn = {9781118525258},
title = {Basics of population genetics: quantifying neutral and adaptive genetic variation for landscape genetic studies},
booktitle = {Landscape Genetics},
chapter = {3},
pages = {35-57},
doi = {https://doi.org/10.1002/9781118525258.ch03},
url = {https://onlinelibrary.wiley.com/doi/abs/10.1002/9781118525258.ch03},
eprint = {https://onlinelibrary.wiley.com/doi/pdf/10.1002/9781118525258.ch03},
year = {2015},
}

@inproceedings{bella21,
  author       = {Bella, G. and Batsuren, K. and Giunchiglia, F.},
  title        = {A database and visualization of the similarity of contemporary lexicons},
  booktitle    = {Proceedings of the 24th International Conference on Text, Speech, and Dialogue, Olomouc, Czech Republik},
  year         = {2021},
  editor       = {Ekstein, K. and Partl, F. and Konopik, M.},
  pages        = {95--104},
  publisher    = {Springer Nature, Switzerland}
}

@article{bowcock94,
issn = {0028-0836},
journal = {Nature},
pages = {455--457},
volume = {368},
publisher = {Nature Publishing},
number = {6470},
year = {1994},
title = {High resolution of human evolutionary trees with polymorphic microsatellites},
copyright = {Copyright 2018 Elsevier B.V., All rights reserved.},
address = {London},
author = {Bowcock, A. M. and Ruiz-Linares, A. and Tomfohrde, J. and Minch, E. and Kidd, J. R. and Cavalli-Sforza, L. L.},
}

@article{bradburd2018,
    author = {Bradburd, G. S. and Coop, G. M. and Ralph, P. L.},
    title = "{Inferring continuous and discrete population genetic structure across space}",
    journal = {Genetics},
    volume = {210},
    number = {1},
    pages = {33-52},
    year = {2018},
    issn = {1943-2631},
    doi = {10.1534/genetics.118.301333},
    url = {https://doi.org/10.1534/genetics.118.301333},
    eprint = {https://academic.oup.com/genetics/article-pdf/210/1/33/49481005/genetics0033.pdf},
}

@article{burbrink2021,
    author = {Burbrink, F. T. and Ruane, S.},
    title = "{Contemporary philosophy and methods for studying speciation and delimiting species}",
    journal = {Ichthyology \& Herpetology},
    volume = {109},
    number = {3},
    pages = {874--894},
    year = {2021},
    issn = {2766-1512},
    doi = {10.1643/h2020073},
    url = {https://doi.org/10.1643/h2020073},
    eprint = {https://meridian.allenpress.com/copeia/article-pdf/109/3/874/2926176/i2766-1520-109-3-874.pdf},
}

@article{carstens2013,
  title={How to fail at species delimitation},
  author={Carstens, B. C. and Pelletier, T. A. and Reid, N. M. and Satler, J. D.},
  journal={Molecular Ecology},
  volume={22},
  number={17},
  pages={4369--4383},
  year={2013},
  publisher={Wiley Online Library}
}

@article{cavallisforza1967,
issn = {0002-9297},
journal = {American Journal of Human Genetics},
pages = {233--257},
volume = {19},
number = {3 Pt 1},
year = {1967},
title = {Phylogenetic analysis. Models and estimation procedures},
language = {eng},
address = {United States},
author = {Cavalli-Sforza, L. L. and Edwards, A. W.},
}

@article{cayuela2018,
issn = {0962-1083},
journal = {Molecular Ecology},
pages = {3976--4010},
volume = {27},
publisher = {Wiley Subscription Services, Inc},
number = {20},
year = {2018},
title = {Demographic and genetic approaches to study dispersal in wild animal populations: a methodological review},
copyright = {2018 John Wiley \& Sons Ltd},
address = {England},
author = {Cayuela, H. and Rougemont, Q. and Prunier, J. G. and Moore, J. and Clobert, J. and Besnard, A. and Bernatchez, L.},
}

@article{clarke2002,
  title={Confidence limits for regression relationships between distance matrices: estimating gene flow with distance},
  author={Clarke, R. T. and Rothery, P. and Raybould, A. F.},
  journal={Journal of Agricultural, Biological, and Environmental Statistics},
  volume={7},
  number={3},
  pages={361--372},
  year={2002},
  publisher={Springer}
}

@article{dequeiroz2007,
    author = {De Queiroz, K.},
    title = "{Species concepts and species delimitation}",
    journal = {Systematic Biology},
    volume = {56},
    number = {6},
    pages = {879--886},
    year = {2007},
    issn = {1063-5157},
    doi = {10.1080/10635150701701083},
    url = {https://doi.org/10.1080/10635150701701083},
    eprint = {https://academic.oup.com/sysbio/article-pdf/56/6/879/24203468/56-6-879.pdf},
}

@article{edwards2014,
  title={Species detection and individual assignment in species delimitation: can integrative data increase efficacy?},
  author={Edwards, D. L. and Knowles, L L.},
  journal={Proceedings of the Royal Society B: Biological Sciences},
  volume={281},
  number={1777},
  pages={20132765},
  year={2014},
  publisher={The Royal Society}
}

@book{efron1993,
publisher = {Chapman \& Hall},
booktitle = {An introduction to the bootstrap},
isbn = {978-04-12-04231-7},
year = {1993},
title = {An introduction to the bootstrap},
address = {New York London},
author = {Efron, B. and Tibshirani, R.},
lccn = {978-04-12-04231-7},
}

@article{frantz2009,
author = {Frantz, A. C. and Cellina, S. and Krier, A. and Schley, L. and Burke, T.},
title = {Using spatial Bayesian methods to determine the genetic structure of a continuously distributed population: clusters or isolation by distance?},
journal = {Journal of Applied Ecology},
volume = {46},
number = {2},
pages = {493-505},
doi = {https://doi.org/10.1111/j.1365-2664.2008.01606.x},
url = {https://besjournals.onlinelibrary.wiley.com/doi/abs/10.1111/j.1365-2664.2008.01606.x},
eprint = {https://besjournals.onlinelibrary.wiley.com/doi/pdf/10.1111/j.1365-2664.2008.01606.x},
year = {2009}
}

@article{genepop,
author = {Rousset, F.},
title = {genepop’007: a complete re-implementation of the genepop software for Windows and Linux},
journal = {Molecular Ecology Resources},
volume = {8},
number = {1},
pages = {103-106},
doi = {https://doi.org/10.1111/j.1471-8286.2007.01931.x},
url = {https://onlinelibrary.wiley.com/doi/abs/10.1111/j.1471-8286.2007.01931.x},
eprint = {https://onlinelibrary.wiley.com/doi/pdf/10.1111/j.1471-8286.2007.01931.x},
year = {2008}
}

@article{goslee2007,
  title={The ecodist package for dissimilarity-based analysis of ecological data},
  author={Goslee, S. C. and Urban, D. L},
  journal={Journal of Statistical Software},
  volume={22},
  pages={1--19},
  year={2007}
}

@article{gratton2016,
issn = {1063-5157},
journal = {Systematic biology},
pages = {292--303},
volume = {65},
publisher = {Oxford University Press},
number = {2},
year = {2016},
title = {Testing classical species properties with contemporary data: how "bad species" in the Brassy Ringlets (Erebia tyndarus complex, Lepidoptera) turned good},
copyright = {Copyright © 2016 Society of Systematic Biologists},
address = {England},
author = {Gratton, P. and Trucchi, E. and Trasatti, A. and Riccarducci, G. and Marta, S. and Allegrucci, G. and Cesaroni, D. and Sbordoni, V.},
}

@article{gspace2021,
  title={GSpace: an exact coalescence simulator of recombining genomes under isolation by distance},
  author={Virgoulay, T. and Rousset, F. and Leblois, R.},
  journal={Bioinformatics},
  volume={37},
  number={20},
  pages={3673--3675},
  year={2021},
  publisher={Oxford University Press}
}

@misc{SLiM_manual,
    title = {SLiM Manual},
    howpublished = {\url{https://messerlab.org/slim/}},
    note = {Accessed: 2023-12-07},
    author = {Haller, B. C. and Messer, P. W.},
    year = {2021},
}

@article{guillot2013,
author = {Guillot, G. and Rousset, F.},
title = {Dismantling the Mantel tests},
journal = {Methods in Ecology and Evolution},
volume = {4},
number = {4},
pages = {336-344},
doi = {https://doi.org/10.1111/2041-210x.12018},
url = {https://besjournals.onlinelibrary.wiley.com/doi/abs/10.1111/2041-210x.12018},
eprint = {https://besjournals.onlinelibrary.wiley.com/doi/pdf/10.1111/2041-210x.12018},
year = {2013}
}

@article{hausdorf2009,
    author = {Hausdorf, B.},
    title = "{Progress toward a general species concept}",
    journal = {Evolution},
    volume = {65},
    number = {4},
    pages = {923-931},
    year = {2011},
    month = {04},
    issn = {0014-3820},
    doi = {10.1111/j.1558-5646.2011.01231.x},
    url = {https://doi.org/10.1111/j.1558-5646.2011.01231.x},
    eprint = {https://academic.oup.com/evolut/article-pdf/65/4/923/47936342/evolut0923.pdf},
}

@article{hh20,
author = {Hausdorf, B. and Hennig, C.},
title = {Species delimitation and geography},
journal = {Molecular Ecology Resources},
volume = {20},
number = {4},
pages = {950-960},
doi = {https://doi.org/10.1111/1755-0998.13184},
url = {https://onlinelibrary.wiley.com/doi/abs/10.1111/1755-0998.13184},
year = {2020}
}

@article{hutchison1999,
issn = {0014-3820},
journal = {Evolution},
pages = {1898--1914},
volume = {53},
publisher = {Society for the Study of Evolution},
number = {6},
year = {1999},
title = {Correlation of pairwise genetic and geographic distance measures: inferring the relative influences of gene flow and drift on the distribution of genetic variability},
copyright = {Copyright 1999 The Society for the Study of Evolution},
address = {United States},
author = {Hutchison, D. W. and Templeton, A. R.},
}

@incollection{jukes1969,
title = {Evolution of protein molecules},
editor = {H. N. Munro},
booktitle = {Mammalian Protein Metabolism},
publisher = {Academic Press},
pages = {21-132},
year = {1969},
isbn = {978-1-4832-3211-9},
doi = {https://doi.org/10.1016/B978-1-4832-3211-9.50009-7},
url = {https://www.sciencedirect.com/science/article/pii/B9781483232119500097},
author = {T. H. Jukes and C. R. Cantor}
}

@article{kimura64,
issn = {0016-6731},
journal = {Genetics},
pages = {561--576},
volume = {49},
number = {4},
year = {1964},
title = {The stepping stone model of population structure and the decrease of genetic correlation with distance},
address = {United States},
author = {Kimura, M. and Weiss, G. H.},
}

@article{ibdsim2009,
  title={IBDSim: a computer program to simulate genotypic data under isolation by distance},
  author={Leblois, R. and Estoup, A. and Rousset, F.},
  journal={Molecular Ecology Resources},
  volume={9},
  number={1},
  pages={107--109},
  year={2009},
  publisher={Wiley Online Library}
}

@article{ishida2009,
 ISSN = {00318248, 1539767X},
 URL = {https://www.jstor.org/stable/10.1086/605802},
 author = {Ishida, Y.},
 journal = {Philosophy of Science},
 number = {5},
 pages = {784--796},
 publisher = {[The University of Chicago Press, Philosophy of Science Association]},
 title = {Sewall Wright and Gustave Malécot on isolation by distance},
 urldate = {2023-01-18},
 volume = {76},
 year = {2009}
}

@article{leache2014,
issn = {1063-5157},
journal = {Systematic biology},
pages = {534--542},
volume = {63},
publisher = {Oxford University Press},
number = {4},
year = {2014},
title = {Species delimitation using genome-wide SNP data},
copyright = {Copyright © 2014 Society of Systematic Biologists},
address = {England},
author = {Leaché, A. D. and Fujita, M. K. and Minin, V. N. and Bouckaert, R. R.},
}

@book{legendre2012,
series = {Developments in environmental modelling ; 24},
publisher = {Elsevier},
isbn = {0-444-53868-2},
year = {2012},
title = {Numerical ecology},
edition = {3rd English ed.},
address = {Amsterdam ; Boston},
author = {Legendre, P. and Legendre, L.},
}

@article{legendre2015,
issn = {2041-210X},
journal = {Methods in ecology and evolution},
pages = {1239--1247},
volume = {6},
publisher = {Wiley},
number = {11},
year = {2015},
title = {Should the Mantel test be used in spatial analysis?},
copyright = {2015 The Authors. Methods in Ecology and Evolution © 2015 British Ecological Society},
address = {HOBOKEN},
author = {Legendre, P. and Fortin, M.-J. and Borcard, D. and Peres‐Neto, P.},
}

@article{legendre2000,
  title={Comparison of permutation methods for the partial correlation and partial Mantel tests},
  author={Legendre, P.},
  journal={Journal of statistical computation and simulation},
  volume={67},
  number={1},
  pages={37--73},
  year={2000},
  publisher={Taylor \& Francis}
}

@article{lme4,
    title = {Fitting Linear Mixed-Effects Models Using {lme4}},
    author = {D. Bates and M. M{\"a}chler and B. Bolker and S. Walker},
    journal = {Journal of Statistical Software},
    year = {2015},
    volume = {67},
    number = {1},
    pages = {1--48},
    doi = {10.18637/jss.v067.i01},
}

@article{mantel67,
    author = {Mantel, N.},
    title = "{The detection of disease clustering and a generalized regression approach}",
    journal = {Cancer Research},
    volume = {27},
    number = {2 Part 1},
    pages = {209-220},
    year = {1967},
    issn = {0008-5472},
    eprint = {https://aacrjournals.org/cancerres/article-pdf/27/2\_Part\_1/209/2382183/cr0272p10209.pdf},
}

@article{mcrae2006,
 ISSN = {00143820, 15585646},
 URL = {http://www.jstor.org/stable/4095372},
 author = {McRae, B. H.},
 journal = {Evolution},
 number = {8},
 pages = {1551--1561},
 publisher = {[Society for the Study of Evolution, Wiley]},
 title = {Isolation by resistance},
 urldate = {2023-05-25},
 volume = {60},
 year = {2006}
}

@article{mcrae2008,
issn = {0012-9658},
journal = {Ecology (Durham)},
pages = {2712--2724},
volume = {89},
publisher = {Ecological Society of America},
number = {10},
year = {2008},
title = {Using circuit theory to model connectivity in ecology, evolution and conservation},
copyright = {Copyright 2008 Ecological Society of America},
address = {Washington, DC},
author = {McRae, B. H. and Dickson, B. G. and Keitt, T. H. and Shah, V. B.},
}

@article{medrano2014,
issn = {1058-5893},
journal = {International journal of plant sciences},
pages = {501--517},
volume = {175},
publisher = {University of Chicago Press},
number = {5},
year = {2014},
title = {Population Genetics Methods Applied to a Species Delimitation Problem: Endemic Trumpet Daffodils (Narcissus Section Pseudonarcissi) from the Southern Iberian Peninsula},
copyright = {2014 by The University of Chicago. All rights reserved.},
address = {CHICAGO},
author = {Medrano, M. and López-Perea, E. and Herrera, C. M.},
}

@article{meirmans2012,
author = {Meirmans, P. G.},
title = {The trouble with isolation by distance},
journal = {Molecular Ecology},
volume = {21},
number = {12},
pages = {2839-2846},
doi = {https://doi.org/10.1111/j.1365-294X.2012.05578.x},
url = {https://onlinelibrary.wiley.com/doi/abs/10.1111/j.1365-294X.2012.05578.x},
eprint = {https://onlinelibrary.wiley.com/doi/pdf/10.1111/j.1365-294X.2012.05578.x},
year = {2012}
}

@article{miller74,
author = {Miller, Rupert G.},
title = {The jackknife--a review},
journal = {Biometrika},
volume = {61},
number = {1},
pages = {1--15},
year = {1974},
doi = {10.1093/biomet/61.1.1}
}

@article{peterman2021,
author = {Peterman, W. E. and Pope, N. S.},
title = {The use and misuse of regression models in landscape genetic analyses},
journal = {Molecular Ecology},
volume = {30},
number = {1},
pages = {37-47},
doi = {https://doi.org/10.1111/mec.15716},
url = {https://onlinelibrary.wiley.com/doi/abs/10.1111/mec.15716},
eprint = {https://onlinelibrary.wiley.com/doi/pdf/10.1111/mec.15716},
year = {2021}
}

@article{pope2015,
author = {Pope, L. C. and Liggins, L. and Keyse, J. and Carvalho, S. B. and Riginos, C.},
title = {Not the time or the place: the missing spatio-temporal link in publicly available genetic data},
journal = {Molecular Ecology},
volume = {24},
number = {15},
pages = {3802-3809},
doi = {https://doi.org/10.1111/mec.13254},
url = {https://onlinelibrary.wiley.com/doi/abs/10.1111/mec.13254},
eprint = {https://onlinelibrary.wiley.com/doi/pdf/10.1111/mec.13254},
year = {2015}
}

@misc{prabclus,
  author = {Hausdorf, B. and Hennig, C.},
  title = {Package 'prabclus', version 2.3-2},
  year = 2019,
  url = {https://CRAN.R-project.org/package=prabclus}
}

@article{pritchard2000,
issn = {0016-6731},
journal = {Genetics (Austin)},
pages = {945--959},
volume = {155},
publisher = {Genetics Soc America},
number = {2},
year = {2000},
title = {Inference of population structure using multilocus genotype data},
copyright = {Copyright 2007 Elsevier B.V., All rights reserved. Medline is the source for the MeSH terms of this document.},
address = {United States},
author = {Pritchard, J. K. and Stephens, M. and Donnelly, P.},
}

@incollection{rannala2020,
author = {Rannala, B. and Yang, Z.},
year = {2020},
publisher={Self published},
title = {Species delimitation},
pages = {5.5:1--5.5:18},
booktitle = {Phylogenetics in the genomic era},
editor = {Scornavacca, C. and Delsuc, F. and Galtier, N.}
}

@article{raxworthy2007,
    author = {Raxworthy, C. J. and Ingram, C. M. and Rabibisoa, N. and Pearson, R. G.},
    title = "{Applications of ecological niche modeling for species delimitation: A review and empirical evaluation using day geckos (Phelsuma) from Madagascar}",
    journal = {Systematic Biology},
    volume = {56},
    number = {6},
    pages = {907-923},
    year = {2007},
    issn = {1063-5157},
    doi = {10.1080/10635150701775111},
    url = {https://doi.org/10.1080/10635150701775111},
    eprint = {https://academic.oup.com/sysbio/article-pdf/56/6/907/26564243/10635150701775111.pdf},
}

@article{resistanceGA,
author = {Peterman, W. E.},
title = {ResistanceGA: An R package for the optimization of resistance surfaces using genetic algorithms},
journal = {Methods in Ecology and Evolution},
volume = {9},
number = {6},
pages = {1638-1647},
doi = {https://doi.org/10.1111/2041-210X.12984},
url = {https://besjournals.onlinelibrary.wiley.com/doi/abs/10.1111/2041-210X.12984},
eprint = {https://besjournals.onlinelibrary.wiley.com/doi/pdf/10.1111/2041-210X.12984},
year = {2018}
}

@article{rissler2007,
    author = {Rissler, L. J. and Apodaca, J. J.},
    title = "{Adding more ecology into species delimitation: ecological niche models and phylogeography help define cryptic species in the black salamander (Aneides flavipunctatus)}",
    journal = {Systematic Biology},
    volume = {56},
    number = {6},
    pages = {924-942},
    year = {2007},
    issn = {1063-5157},
    doi = {10.1080/10635150701703063},
    url = {https://doi.org/10.1080/10635150701703063},
    eprint = {https://academic.oup.com/sysbio/article-pdf/56/6/924/24204319/56-6-924.pdf},
}

@article{rousset97,
    author = {Rousset, F.},
    title = "{Genetic differentiation and estimation of gene flow from F-statistics under isolation by distance}",
    journal = {Genetics},
    volume = {145},
    number = {4},
    pages = {1219-1228},
    year = {1997},
    issn = {1943-2631},
    doi = {10.1093/genetics/145.4.1219},
}

@article{royston2007,
  title={Profile likelihood for estimation and confidence intervals},
  author={Royston, P.},
  journal={The Stata Journal},
  volume={7},
  number={3},
  pages={376--387},
  year={2007},
  publisher={SAGE Publications Sage CA: Los Angeles, CA}
}

@misc{SAS,
    title = {SAS/STAT user’s guide. Ver. 8},
    author = {{SAS Institute}},
    year = {2001},
    publisher = {SAS Institute, Inc.},
    address = {Cary, NC},
}

@article{scapini2002,
  title={Multiple regression analysis of the sources of variation in orientation of two sympatric sandhoppers, Talitrus saltator and Talorchestia brito, from an exposed Mediterranean beach},
  author={Scapini, F. and Aloia, A. and Bouslama, M. F. and Chelazzi, L. and Colombini, I. and ElGtari, M. and Fallaci, M. and Marchetti, G. M.},
  journal={Behavioral Ecology and Sociobiology},
  volume={51},
  pages={403--414},
  year={2002},
  publisher={Springer}
}

@article{shirk2017,
author = {Shirk, A. J. and Landguth, E. L. and Cushman, S. A.},
title = {A comparison of regression methods for model selection in individual-based landscape genetic analysis},
journal = {Molecular Ecology Resources},
volume = {18},
number = {1},
pages = {55-67},
doi = {https://doi.org/10.1111/1755-0998.12709},
url = {https://onlinelibrary.wiley.com/doi/abs/10.1111/1755-0998.12709},
eprint = {https://onlinelibrary.wiley.com/doi/pdf/10.1111/1755-0998.12709},
year = {2018}
}

@article{slatkin93,
issn = {0014-3820},
journal = {Evolution},
pages = {264--279},
volume = {47},
publisher = {Society for the Study of Evolution},
number = {1},
year = {1993},
title = {Isolation by distance in equilibrium and non-equilibrium populations},
copyright = {The Society for the Study of Evolution},
address = {Malden, MA},
author = {Slatkin, M.},
}

@article{slim2023,
  title={SLiM 4: multispecies eco-evolutionary modeling},
  author={Haller, B. C. and Messer, P. W.},
  journal={The American Naturalist},
  volume={201},
  number={5},
  pages={E127--E139},
  year={2023},
  publisher={The University of Chicago Press Chicago, IL}
}

@article{SLS86,
 ISSN = {00397989},
 URL = {http://www.jstor.org/stable/2413122},
 author = {P. E. Smouse and J. C. Long and R. R. Sokal},
 journal = {Systematic Zoology},
 number = {4},
 pages = {627--632},
 publisher = {Oxford University Press, Society of Systematic Biologists, Taylor \& Francis, Ltd.},
 title = {Multiple Regression and Correlation Extensions of the Mantel Test of Matrix Correspondence},
 urldate = {2023-07-24},
 volume = {35},
 year = {1986}
}

@article{spriggs2019,
    author = {Spriggs, E. L. and Eaton, D. A. R. and Sweeney, P. W. and Schlutius, C. and Edwards, E. J. and Donoghue, M. J.},
    title = "{Restriction-site-associated DNA sequencing reveals a cryptic Viburnum species on the north American coastal plain}",
    journal = {Systematic Biology},
    volume = {68},
    number = {2},
    pages = {187-203},
    year = {2018},
    issn = {1063-5157},
    doi = {10.1093/sysbio/syy084},
    url = {https://doi.org/10.1093/sysbio/syy084},
    eprint = {https://academic.oup.com/sysbio/article-pdf/68/2/187/27739208/syy084.pdf},
}

@article{storfer2010,
  title={Landscape genetics: where are we now?},
  author={Storfer, A. and Murphy, M. A. and Spear, S. F. and Holderegger, R. and Waits, L. P.},
  journal={Molecular Ecology},
  volume={19},
  number={17},
  pages={3496--3514},
  year={2010},
  publisher={Wiley Online Library}
}

@article{szekely2007,
author = {G. J. Sz{\'e}kely and M. L. Rizzo and N. K. Bakirov},
title = {{Measuring and testing dependence by correlation of distances}},
volume = {35},
journal = {The Annals of Statistics},
number = {6},
publisher = {Institute of Mathematical Statistics},
pages = {2769 -- 2794},
year = {2007},
doi = {10.1214/009053607000000505},
URL = {https://doi.org/10.1214/009053607000000505}
}

@article{vanstrien2012,
issn = {0962-1083},
journal = {Molecular Ecology},
pages = {4010--4023},
volume = {21},
publisher = {Blackwell Publishing Ltd},
number = {16},
year = {2012},
title = {A new analytical approach to landscape genetic modelling: least-cost transect analysis and linear mixed models},
edition = {Received 6 July 2011; revision received 2 May 2012; accepted 16 May 2012},
copyright = {2012 Blackwell Publishing Ltd},
address = {Oxford, UK},
author = {Van Strien, M. J. and Keller, D. and Holderegger, R.},
}

@article{vekemanshardy2004,
author = {Vekemans, X. and Hardy, O. J.},
title = {New insights from fine-scale spatial genetic structure analyses in plant populations},
journal = {Molecular Ecology},
volume = {13},
number = {4},
pages = {921-935},
doi = {https://doi.org/10.1046/j.1365-294X.2004.02076.x},
url = {https://onlinelibrary.wiley.com/doi/abs/10.1046/j.1365-294X.2004.02076.x},
eprint = {https://onlinelibrary.wiley.com/doi/pdf/10.1046/j.1365-294X.2004.02076.x},
year = {2004}
}

@article{venzon1988,
  title={A method for computing profile-likelihood-based confidence intervals},
  author={Venzon, D. J. and Moolgavkar, S. H.},
  journal={Journal of the Royal Statistical Society: Series C (Applied Statistics)},
  volume={37},
  number={1},
  pages={87--94},
  year={1988},
  publisher={Wiley Online Library}
}

@article{weir1984,
issn = {0014-3820},
journal = {Evolution},
pages = {1358--1370},
volume = {38},
publisher = {Society for the Study of Evolution},
number = {6},
year = {1984},
title = {Estimating F-statistics for the analysis of population structure},
copyright = {The Society for the Study of Evolution},
address = {Malden, MA},
author = {Weir, B. S. and Cockerham, C. Clark},
}

@article{welch47,
    author = {Welch, B. L.},
    title = "{The generalization of "Student's" problem when several different population variances are involved}",
    journal = {Biometrika},
    volume = {34},
    number = {1-2},
    pages = {28-35},
    year = {1947},
    issn = {0006-3444},
    doi = {10.1093/biomet/34.1-2.28},
    url = {https://doi.org/10.1093/biomet/34.1-2.28},
    eprint = {https://academic.oup.com/biomet/article-pdf/34/1-2/28/553093/34-1-2-28.pdf},
}

@book{west2022,
  title={Linear mixed models: a practical guide using statistical software},
  author={West, B. T. and Welch, K. B. and Galecki, A. T.},
  year={2022},
  publisher={Crc Press}
}

@article{yang2004,
author = {R. Yang},
title = {A Likelihood-based approach to estimating and testing for isolation by distance},
journal = {Evolution},
volume = {58},
number = {8},
pages = {1839-1845},
doi = {https://doi.org/10.1111/j.0014-3820.2004.tb00466.x},
url = {https://onlinelibrary.wiley.com/doi/abs/10.1111/j.0014-3820.2004.tb00466.x},
eprint = {https://onlinelibrary.wiley.com/doi/pdf/10.1111/j.0014-3820.2004.tb00466.x},
year = {2004}
}

@article{shaowu89,
author = {Jun Shao and C. F. J. Wu},
title = {{A General Theory for Jackknife Variance Estimation}},
volume = {17},
journal = {The Annals of Statistics},
number = {3},
publisher = {Institute of Mathematical Statistics},
pages = {1176 -- 1197},
year = {1989},
doi = {10.1214/aos/1176347263},
URL = {https://doi.org/10.1214/aos/1176347263}
}

@article{dietz83,
 ISSN = {00397989},
 URL = {http://www.jstor.org/stable/2413216},
 author = {E. Jacquelin Dietz},
 journal = {Systematic Zoology},
 number = {1},
 pages = {21--26},
 publisher = {[Oxford University Press, Society of Systematic Biologists, Taylor & Francis, Ltd.]},
 title = {Permutation Tests for Association between Two Distance Matrices},
 urldate = {2025-07-17},
 volume = {32},
 year = {1983}
}

\appendix
\section{Software code and distance-distance plots}
\label{appendix:code}

\subsection{SLiM}
Script editing, software executions and data analysis were carried out in \texttt{R}.
The variables defined with the \texttt{defineConstant}  command were set at each simulation iteration before feeding the script to the SLiM executable: for instance, \texttt{nsam} would take value $6$, $15$ or $45$ according to the scenario at hand.

\lstset{language=C++, otherkeywords={p0, m1, g1}, deletekeywords={line, info, pos, rnorm, sample, count, paste, float}, morecomment=[f][\color{OliveGreen}][0]{//}}

\begin{lstlisting}[caption = {SLiM script for the \textit{split} scenario}]
initialize() {
initializeSLiMOptions(keepPedigrees = T, dimensionality="xy", nucleotideBased=T);
	defineConstant("L", 1e3);
	initializeAncestralNucleotides(randomNucleotides(L));
	initializeMutationTypeNuc("m1", 0.5, "f", 0.0);
	initializeGenomicElementType("g1", m1, 1.0, mmJukesCantor(2.5e-3));
	initializeGenomicElement(g1, 0, L-1);
	initializeRecombinationRate(rates = 1e-8);
	
	defineConstant(symbol="nsim", value=200);
	defineConstant(symbol="nsam", value=45);
	defineConstant(symbol="i1sd", value=1);
	defineConstant(symbol="i2maxd", value=3);
	defineConstant(symbol="nneigh", value=3);
	defineConstant(symbol="childsd", value=9);
	
	// spatial competition
	initializeInteractionType(1, "xy", reciprocal=T, maxDistance=3*i1sd);
		i1.setInteractionFunction("n", i1sd/2, i1sd);
	
	
	// spatial mate choice
	initializeInteractionType(2, "xy", reciprocal=T, maxDistance=i2maxd);
}

1 late() {
	sim.addSubpop("p0", nsim*2);
	p0.setSpatialBounds(c(50.00, 50.00, 150.00, 150.00));
	p0.individuals.setSpatialPosition(p0.pointUniform(nsim*2));
}

1:100 late() {
	i1.evaluate(p0);
	inds = p0.individuals;
	competition = i1.totalOfNeighborStrengths(inds) / size(inds);
	competition = pmin(competition, 0.99);
	inds.fitnessScaling = 1 - competition;
}

2:100 first() {
	i2.evaluate(p0);
}

mateChoice(p0) {
	// nearest-neighbor mate choice
	neighbors = i2.nearestNeighbors(individual, count = nneigh);
	return (size(neighbors) ? sample(neighbors, 1) else float(0));
}
modifyChild(p0) {
	do pos = parent1.spatialPosition + rnorm(2, 0, childsd);
	while (!p0.pointInBounds(pos));
	child.setSpatialPosition(pos);
	
	return T;
}

100 late() { // last generation
	sampledIndividuals = p0.sampleIndividuals(nsam*2);

	out = paste("subpopulation", "pedigreeID", "x", "y", "genome1", "genome2");
	for (i in sampledIndividuals){
		info = paste(i.subpopulation, i.pedigreeID, i.spatialPosition, i.genome1.nucleotides(), i.genome2.nucleotides());
		out = c(out, info);
	}
	writeFile("coordgenomes.txt", out);
}
\end{lstlisting}

\begin{lstlisting}[caption = {SLiM script for the \textit{conspecificity} scenario}]
initialize() {
    initializeSLiMOptions(keepPedigrees = T, dimensionality="xy", nucleotideBased=T);
    defineConstant("L", 1e3);
    initializeAncestralNucleotides(randomNucleotides(L));
    initializeMutationTypeNuc("m1", 0.5, "f", 0.0);
    initializeGenomicElementType("g1", m1, 1.0, mmJukesCantor(2.5e-3));
    initializeGenomicElement(g1, 0, L-1);
    initializeRecombinationRate(rates = 1e-8);
    
    defineConstant(symbol="nsim", value=200);
    defineConstant(symbol="nsam", value=45);
    defineConstant(symbol="i1sd", value=1);
    defineConstant(symbol="i2maxd", value=3);
    defineConstant(symbol="nneigh", value=3);
    defineConstant(symbol="childsd", value=9);
    defineConstant(symbol="separation", value=F);
	
    // spatial competition
    initializeInteractionType(1, "xy", reciprocal=T, maxDistance=3*i1sd);
    i1.setInteractionFunction("n", i1sd/2, i1sd);
	
    // spatial mate choice
    initializeInteractionType(2, "xy", reciprocal=T, maxDistance=i2maxd);
}
1 late() {
    sim.addSubpop("p0", nsim*2);
    p0.setSpatialBounds(c(50.00, 50.00, 150.00, 150.00));
    p0.individuals.setSpatialPosition(p0.pointUniform(nsim*2));
}

1:100 late() {
    i1.evaluate(p0);
    inds = p0.individuals;
    competition = i1.totalOfNeighborStrengths(inds) / size(inds);
    competition = pmin(competition, 0.99);
    inds.fitnessScaling = 1 - competition;
}
2:100 first() {	i2.evaluate(p0); }

90 early() {
    sim.addSubpopSplit("p1", nsim, p0);
    sim.addSubpopSplit("p2", nsim, p0);
    p1.setSpatialBounds(c(50.00, 50.00, 150.00, 150.00));
    p2.setSpatialBounds(c(50.00, 50.00, 150.00, 150.00));
    p1.individuals.setSpatialPosition(p1.pointUniform(nsim));
    p2.individuals.setSpatialPosition(p2.pointUniform(nsim));
}

91: late() {
    i1.evaluate(p1);
    inds = p1.individuals;
    competition = i1.totalOfNeighborStrengths(inds) / size(inds);
    competition = pmin(competition, 0.99);
    inds.fitnessScaling = 1 - competition;
    
    i1.evaluate(p2);
    inds = p2.individuals;
    competition = i1.totalOfNeighborStrengths(inds) / size(inds);
    competition = pmin(competition, 0.99);
    inds.fitnessScaling = 1 - competition;
}

92: first() {
    i2.evaluate(p1);
    i2.evaluate(p2);
}

101 early() { p0.setSubpopulationSize(0); }


101:150 early() {
    if(separation){
p1.setSpatialBounds(p1.spatialBounds + c(0, 0, -1, -1));
p2.setSpatialBounds(p2.spatialBounds + c(1, 1, 0, 0));
    }
}

92: early() {
    migrationProgress = runif(1, min=0.8, max=1);
    p1.setMigrationRates(p2, 0.5 * migrationProgress);
    p2.setMigrationRates(p1, 0.5 * migrationProgress);
}


// NEAREST NEIGHBORS MATE CHOICE
2:91 mateChoice(p0) {
    // nearest-neighbor mate choice
    neighbors = i2.nearestNeighbors(individual, count = nneigh);
    return (size(neighbors) ? sample(neighbors, 1) else float(0));
}
2:91 modifyChild(p0) {
    do pos = parent1.spatialPosition + rnorm(2, 0, childsd);
    while (!p0.pointInBounds(pos));
    child.setSpatialPosition(pos);
    return T;
}

92: mateChoice(p1) {
    // nearest-neighbor mate choice
    neighbors = i2.nearestNeighbors(individual, count = nneigh);
    return (size(neighbors) ? sample(neighbors, 1) else float(0));
}
92: mateChoice(p2) {
    // nearest-neighbor mate choice
    neighbors = i2.nearestNeighbors(individual, count = nneigh);
    return (size(neighbors) ? sample(neighbors, 1) else float(0));
}

92: modifyChild(p1) {
    counter = 1;
    do{
        pos = parent1.spatialPosition + rnorm(2, ifelse(separation, -0.5, 0.0), childsd);
        counter = counter + 1;
    }
    while (!p1.pointInBounds(pos) & counter < 100);
    child.setSpatialPosition(pos);
    return T;
}
92: modifyChild(p2) {
    counter = 1;
    do{
        pos = parent1.spatialPosition + rnorm(2, ifelse(separation, 0.5, 0.0), childsd);
        counter = counter + 1;
    }
    while (!p2.pointInBounds(pos) & counter < 100);
    child.setSpatialPosition(pos);
    return T;
}
150 late() { // last generation
        allind1 = p1.individuals[p1.pointInBounds(p1.individuals.spatialPosition)];
        allind2 = p2.individuals[p2.pointInBounds(p2.individuals.spatialPosition)];
    sampledIndividuals = c(sample(allind1, nsam), sample(allind2, nsam));
    sampledIndividuals.genomes.outputVCF(filePath="tmp.VCF", outputMultiallelics = F, simplifyNucleotides=T);
    
    out = paste("subpopulation", "pedigreeID", "x", "y", "genome1", "genome2");
    for (i in sampledIndividuals){
        info = paste(i.subpopulation, i.pedigreeID, i.spatialPosition, i.genome1.nucleotides(), i.genome2.nucleotides());
	out = c(out, info);
    }
    writeFile("coordgenomes.txt", out);
}
\end{lstlisting}

\begin{lstlisting}[caption = {SLiM script for the \textit{distinctness} scenario}]
species all initialize() {
    defineConstant("L", 1e3);
    defineConstant(symbol="nsim", value=200);
    defineConstant(symbol="nsam", value=45);
    defineConstant(symbol="i1sd", value=1);
    defineConstant(symbol="i2maxd", value=3);
    defineConstant(symbol="nneigh", value=3);
    defineConstant(symbol="childsd", value=9);
	
    // spatial competition
    initializeInteractionType(1, "xy", reciprocal=T, maxDistance=3*i1sd);
    i1.setInteractionFunction("n", i1sd/2, i1sd);
    
    // spatial mate choice
    initializeInteractionType(2, "xy", reciprocal=T, maxDistance=i2maxd);
}

species sunflower initialize() {
    initializeSpecies(tickModulo=1, tickPhase=1, avatar="S");
    initializeSLiMOptions(keepPedigrees = T, dimensionality="xy", nucleotideBased=T);
    initializeAncestralNucleotides(randomNucleotides(L, c(1, 1, 1, 1)));
    initializeMutationTypeNuc("m1", 0.5, "f", 0.0);
    initializeGenomicElementType("g1", m1, 1.0, mmJukesCantor(2.5e-3));
    initializeGenomicElement(g1, 0, L-1);
    initializeRecombinationRate(rates = 1e-8);
}

species tulip initialize() {
    initializeSpecies(tickModulo=1, tickPhase=1, avatar="T");
    initializeSLiMOptions(keepPedigrees = T, dimensionality="xy", nucleotideBased=T);
    initializeAncestralNucleotides(randomNucleotides(L, c(1, 1, 1, 1)));
    initializeMutationTypeNuc("m2", 0.5, "f", 0.0);
    initializeGenomicElementType("g2", m2, 1.0, mmJukesCantor(2.5e-3));
    initializeGenomicElement(g2, 0, L-1);
    initializeRecombinationRate(rates = 1e-8);
}

ticks all 1 early() {
    sunflower.addSubpop("p1", nsim);
    tulip.addSubpop("p2", nsim);
    p1.setSpatialBounds(c(50.00, 50.00, 100.00, 100.00));
    p2.setSpatialBounds(c(100.00, 100.00, 150.00, 150.00));
    p1.individuals.setSpatialPosition(p1.pointUniform(nsim));
    p2.individuals.setSpatialPosition(p2.pointUniform(nsim));
}

ticks all 1: late() {
    i1.evaluate(p1);
    inds = p1.individuals;
    competition = i1.totalOfNeighborStrengths(inds) / size(inds);
    competition = pmin(competition, 0.99);
    inds.fitnessScaling = 1 - competition;
    
    i1.evaluate(p2);
    inds = p2.individuals;
    competition = i1.totalOfNeighborStrengths(inds) / size(inds);
    competition = pmin(competition, 0.99);
    inds.fitnessScaling = 1 - competition;
}

ticks all 2: first() {
	i2.evaluate(p1);
	i2.evaluate(p2);
}

// NEAREST NEIGHBORS MATE CHOICE
species sunflower 2: mateChoice(p1) {
    // nearest-neighbor mate choice
    neighbors = i2.nearestNeighbors(individual, count = nneigh);
    return (size(neighbors) ? sample(neighbors, 1) else float(0));
}
species tulip 2: mateChoice(p2) {
    // nearest-neighbor mate choice
    neighbors = i2.nearestNeighbors(individual, count = nneigh);
    return (size(neighbors) ? sample(neighbors, 1) else float(0));
}
species sunflower modifyChild(p1) {
    do pos = parent1.spatialPosition + rnorm(2, 0, childsd);
    while (!p1.pointInBounds(pos));
    child.setSpatialPosition(pos);
    
    return T;
}
species tulip modifyChild(p2) {
    do pos = parent1.spatialPosition + rnorm(2, 0, childsd);
    while (!p2.pointInBounds(pos));
    child.setSpatialPosition(pos);
    
    return T;
}

ticks all 50 late() { // last generation
    allind1 = p1.individuals;
    allind2 = p2.individuals;
    sampled1 = sample(allind1, nsam);
    sampled2 = sample(allind2, nsam);
    sampledIndividuals = c(sampled1, sampled2);
    
    out = paste("subpopulation", "pedigreeID", "x", "y", "genome1", "genome2");
    for (i in sampledIndividuals){
        info = paste(i.subpopulation, i.pedigreeID, i.spatialPosition, i.genome1.nucleotides(), i.genome2.nucleotides());
        out = c(out, info);
	}
    writeFile("coordgenomes.txt", out);
}
\end{lstlisting}

\begin{figure}[h] 
    \centering
    \includegraphics[width=1\textwidth]{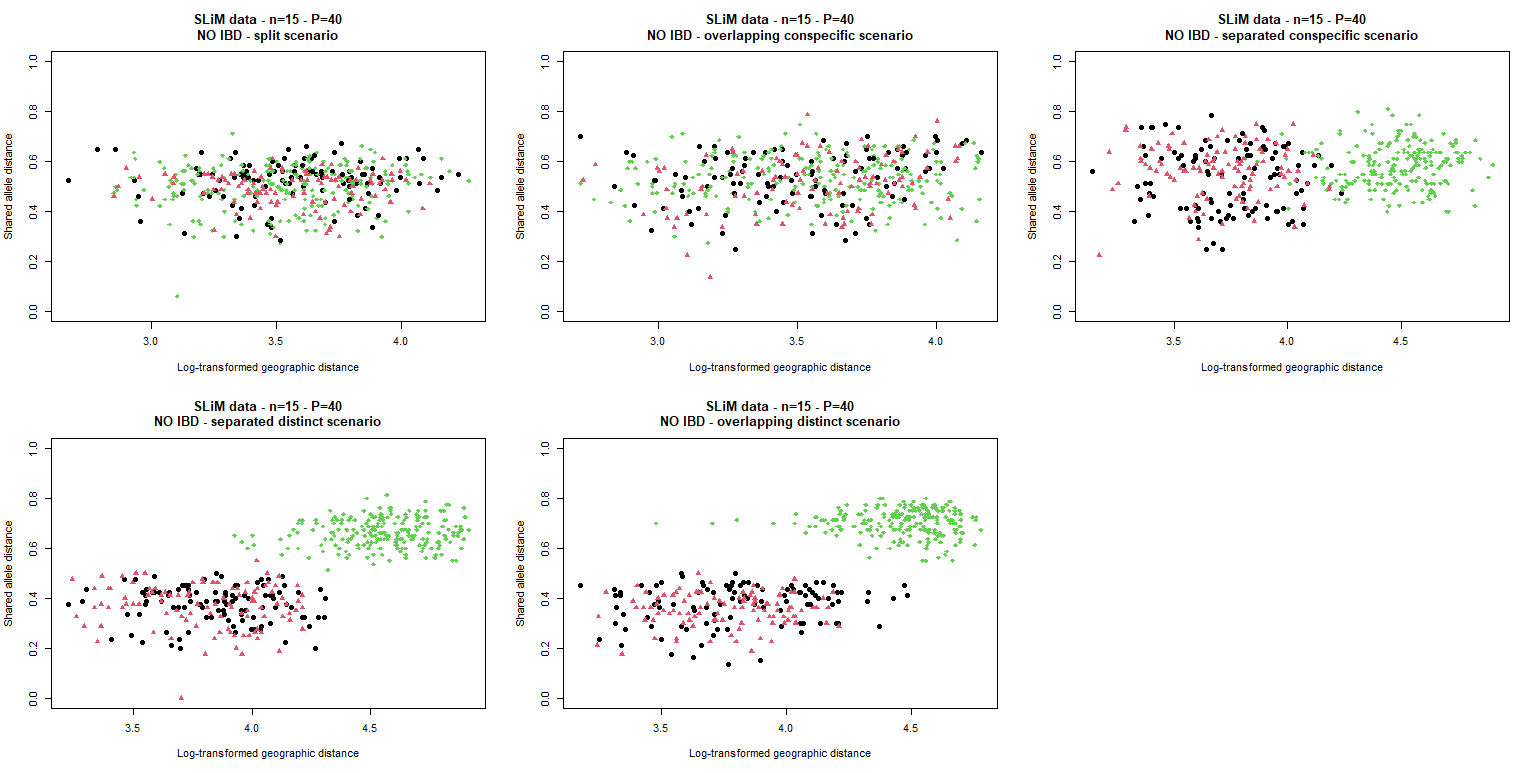}
    \captionsetup{width=.9\linewidth}
    \caption{Distance-distance plots (shared allele distance against log-transformed geographic distance) from the five scenarios simulated with SLiM, \textbf{without IBD behaviour}, $15$ individuals per group and $40$ available loci. Each scenario is specified in the plot title. In black the dissimilarities among individuals belonging to the first group, in red those for the second group and in green the dissimilarities among individuals belonging to different groups.}
    \label{fig:SLiMscenariosNOIBD}
\end{figure}

In Figure \ref{fig:SLiMscenariosNOIBD}, the shared allele distance is plotted against the log-transformed geographic distance for a random dataset from each of the five scenarios simulated by SLiM without IBD behaviour. It is easy to see that, by colour, there was no association between genetic and geographic dissimilarities. The \textit{split} and the \textit{overlapping conspecific} scenarios are similar, with the latter displaying a larger variability in the genetic dissimilarities. In the third scenario, the geographic separation can be clearly seen by the range of the green-colored points. In the multispecies scenarios (bottom row), the range of the between-group genetic dissimilarities was always above that of the within-group ones, supporting distinctness: models that fit an update in the intercept to a linear regression of $\mathbf{D}_y$ against $\mathbf{D}_x$ easily caught this trend in the data.

\begin{figure}[h] 
    \centering
    \includegraphics[width=1\textwidth]{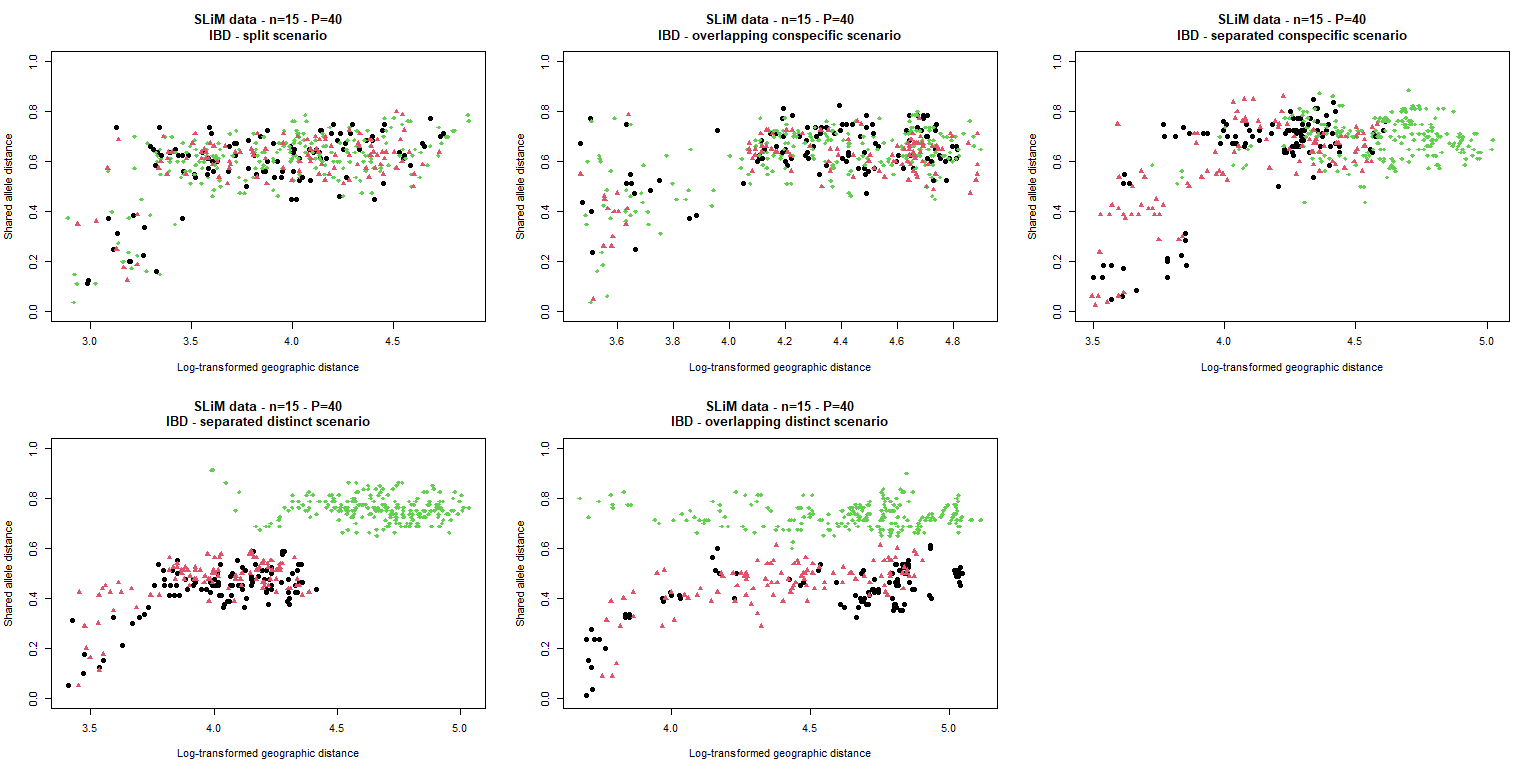}
    \captionsetup{width=.9\linewidth}
    \caption{Distance-distance plots (shared allele distance against log-transformed geographic distance) from the five scenarios simulated with SLiM, \textbf{with IBD behaviour}, $15$ individuals per group and $40$ available loci. Each scenario is specified in the plot title. In black the dissimilarities among individuals belonging to the first group, in red those for the second group and in green the dissimilarities among individuals belonging to different groups.}
    \label{fig:SLiMscenariosIBD}
\end{figure}

In Figure \ref{fig:SLiMscenariosIBD}, the same kind of plots are shown for IBD species. It is apparent how the shared allele distance is positively correlated with geographic distance as far as individuals close in space are considered, but this trend soon plateaus and dissimilarities never take values larger than around $0.9$ (also in the quasi-panmictic datasets). This saturation causes a convexity in the relationship that is not fixed by the logarithmic transformation of the geographic distances. Also in this chart it is easy to notice: a) how within and between-group dissimilarities are separated in the \textit{distinct} scenarios and b) how the IBD behaviour is common to both groups.

\clearpage

\subsection{GSpace}

Script editing, software executions and data analysis were carried out in \texttt{R}.
Variables like \texttt{Sequence\_Size} or \texttt{Dispersal\_Distribution} were modified according to the scenario at hand.
\newline
When only one software execution was involved (trivial conspecificity scenario), all individuals from the two groups were simulated at once, using a longer sequence of random coordinates next to the \texttt{Sample\_Coordinate} variables. In all other situations, the code was run twice with half that number of coordinate pairs, each time fulfilling the geographic separation explained in section \ref{GSpace_section}: e.g., individuals from the first group would have coordinates bound within point $(70, 70)$ and point $(90, 90)$.

\begin{lstlisting}[caption = {Common structure of the GSpace script}]
%%%%%%%% SIMULATION SETTINGS %%%%%%%%%%%%%%%
Setting_Filename = GSpaceSettings.txt
Random_seeds = 11000
Run_Number = 1

%%%%%%%% OUTPUT FILE FORMAT SETTINGS %%%%%%%
Output_Dir = .
Data_File_Name = trial
Data_File_Extension = .txt

Genepop = True
Genepop_ind_file = F
Genepop_Group_All_Samples = T

Approximate_time = F

%%%%%%%% MARKERS SETTINGS %%%%%%%%%%%%%%%%%%
Ploidy = Diploid
Chromosome_number = 1
Mutation_Rate = 0.005
Mutation_Model = KAM
Allelic_Lower_Bound = 240
Allelic_Upper_Bound = 241
Sequence_Size = 100

%%%%%%%% RECOMBINATION SETTINGS %%%%%%%%%%%%
Recombination_Rate = 0.005

%%%%%%%% DEMOGRAPHIC SETTINGS %%%%%%%%%%%%%%
%% LATTICE
Min_Sample_Coord_X = 50
Min_Sample_Coord_Y = 50
Lattice_Size_X = 200
Lattice_Size_Y = 200
Ind_Per_Node = 1

%% DISPERSAL
Dispersal_Distribution = p
Pareto_Shape = 5

Edge_Effects = circular
Total_Emigration_Rate = 0.5
Disp_Dist_Max = 200, 200

%%%%%%%% SAMPLE SETTINGS %%%%%%%%%%%%%%%%%%%
Sample_Coordinate_X = 82,84,70,82,75,71,115,127,130,122,128,124
Sample_Coordinate_Y = 89,76,89,72,73,70,111,116,128,120,117,129

% STATS
Dist_Class_Nbr = 1
Ind_Per_Node_Sampled = 1
Pause = Never

\end{lstlisting}

\begin{figure}[h] 
    \centering
    \includegraphics[width=1\textwidth]{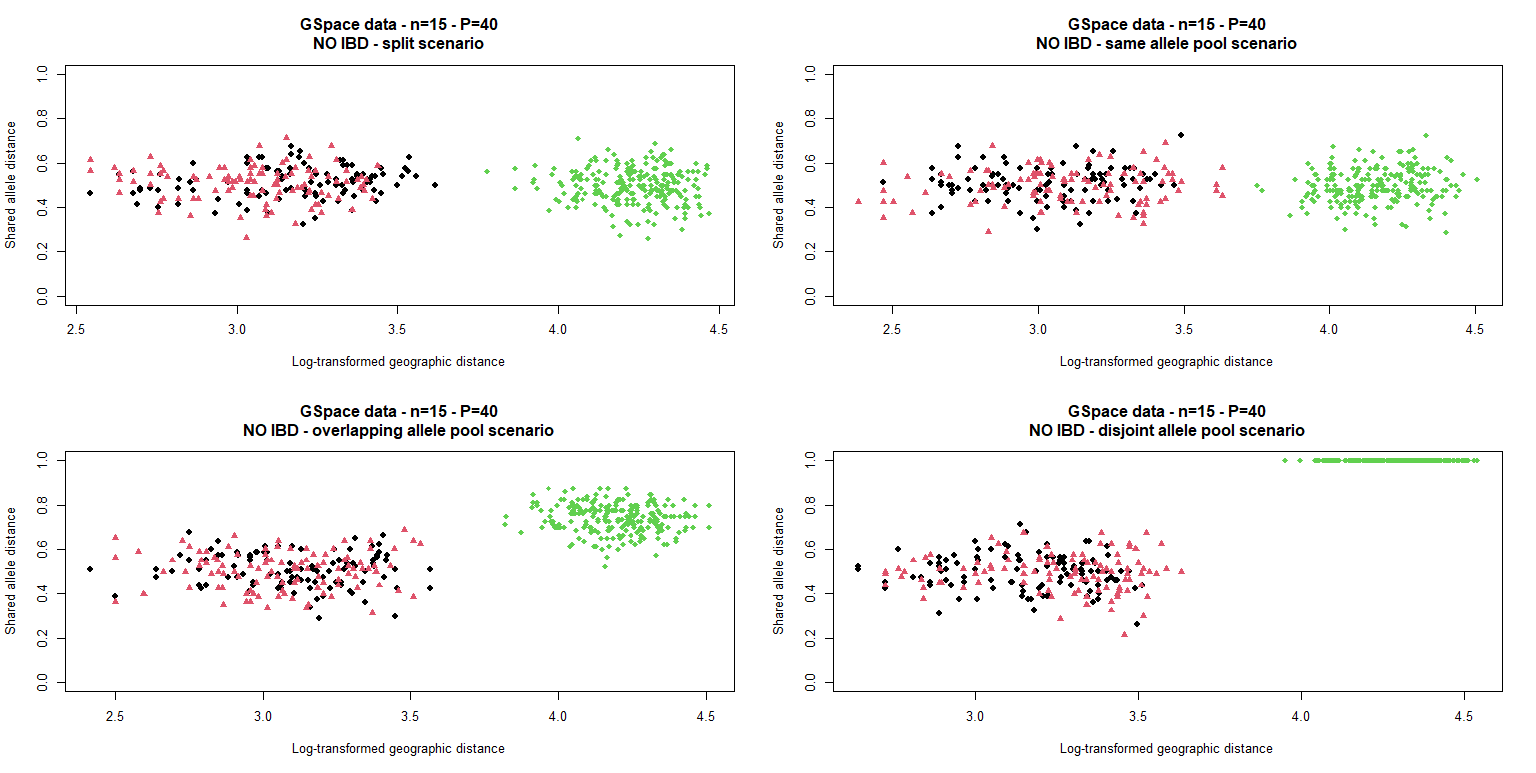}
    \captionsetup{width=.9\linewidth}
    \caption{Distance-distance plots (shared allele distance against log-transformed geographic distance) from the four scenarios simulated with GSpace, \textbf{without IBD behaviour}, $15$ individuals per group and $40$ available loci. Each scenario is specified in the plot title. In black the dissimilarities among individuals belonging to the first group, in red those for the second group and in green the dissimilarities among individuals belonging to different groups.}
    \label{fig:GSpacescenariosNOIBD}
\end{figure}

\begin{figure}[h] 
    \centering
    \includegraphics[width=1\textwidth]{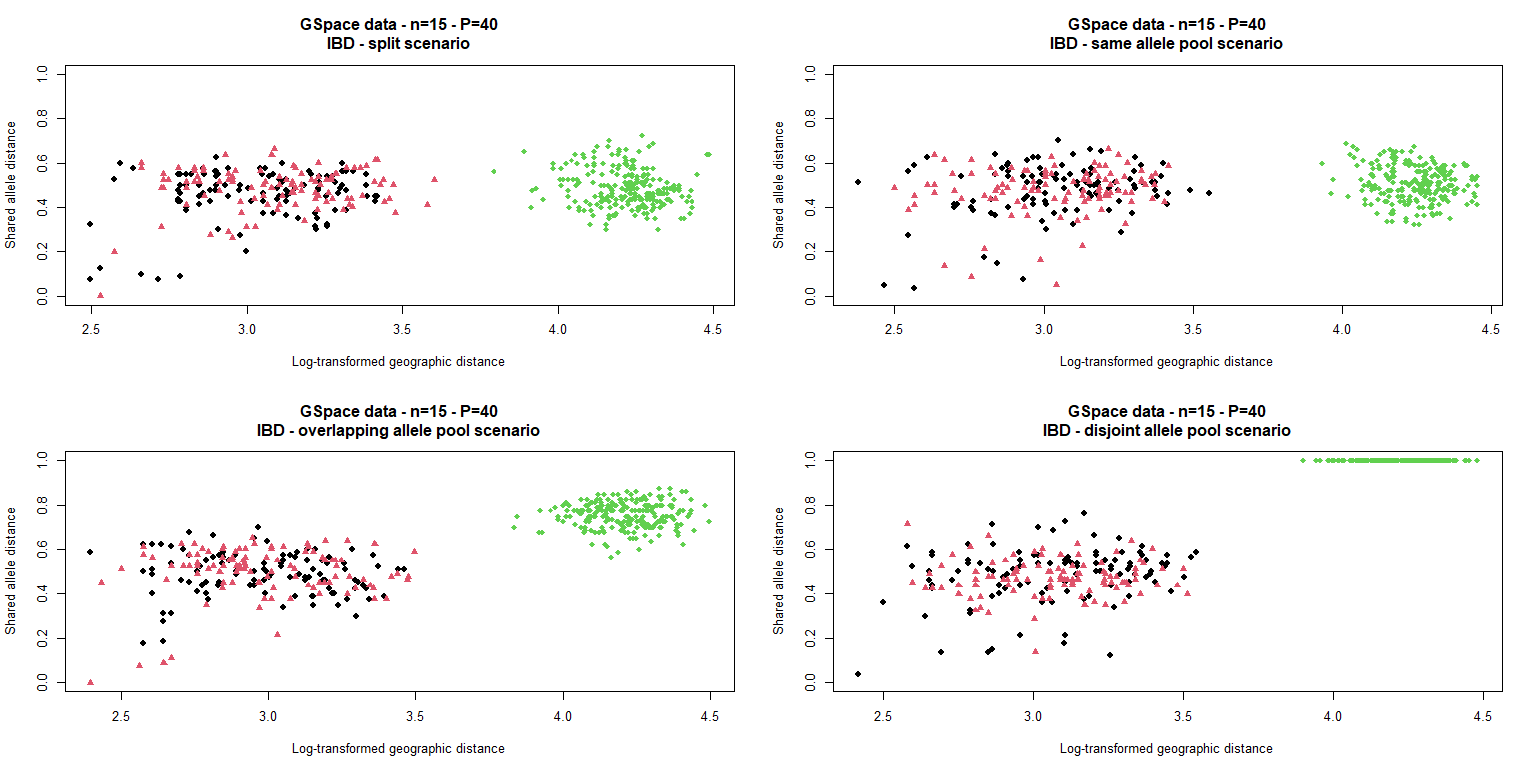}
    \captionsetup{width=.9\linewidth}
    \caption{Distance-distance plots (shared allele distance against log-transformed geographic distance) from the four scenarios simulated with GSpace, \textbf{with IBD behaviour}, $15$ individuals per group and $40$ available loci. Each scenario is specified in the plot title. In black the dissimilarities among individuals belonging to the first group, in red those for the second group and in green the dissimilarities among individuals belonging to different groups.}
    \label{fig:GSpacepowerIBD}
\end{figure}

In Figure \ref{fig:GSpacescenariosNOIBD}, the shared allele distance is plotted against the log-transformed geographic distance for a random dataset from each of the four scenarios simulated by GSpace without IBD behaviour. As expected, there is no association between genetic and geographic dissimilarities within the groups. The \textit{split} and \textit{same allele pool} scenarios look rather similar and a major difference with respect to SLiM distinctness setups lies in the range of the dissimilarities among individuals belonging to different groups: in the \textit{disjoint allele pool} scenario, since the genetic make-ups in the two groups are completely different, all genetic dissimilarities take value $1$.
\newline
In Figure \ref{fig:GSpacepowerIBD}, the same visualizations are provided for species displaying a positive association between genetic and geographic dissimilarities. This relationship is however harder to spot with respect to SLiM simulations as it seems to relate only a small share of the individuals in the two groups. In both Figures, the between-group genetic dissimilarities in the \textit{overlapping allele pool} scenario are centered around the value $0.75$, whereas in all scenarios the bulk of the within-group shared allele distances is centered at $0.5$. This vertical shift in the genetic dissimilarities, which is even stronger in the last scenario, could be quite effectively captured by an update in the intercept of the linear models fitted on this data.

\end{document}